\documentclass[11pt,a4paper]{article}
\usepackage{jheppubmod}
\usepackage{cite}
\usepackage{cancel}
\usepackage[export]{adjustbox}
\usepackage{xcolor}
\usepackage{cite}
\usepackage{relsize}
\usepackage{physics}
\usepackage{psfrag}
\usepackage{cancel}
\usepackage{array}
\usepackage{amssymb}
\usepackage{amsmath}
\usepackage{amsthm}
\usepackage{float}
\usepackage{tikz}
\usetikzlibrary{decorations.markings}
\usepackage{tikz,lipsum,lmodern}
\usepackage[most]{tcolorbox}
\usepackage{xcolor}
\usepackage{epsfig,amsfonts,amssymb,mathtools}
\usepackage{amsmath}
\usepackage{wrapfig,lipsum,booktabs}
\usepackage{subcaption}
\usepackage{graphicx}
\usepackage{framed}

\hypersetup{
	colorlinks=true,
	urlcolor=midnightblue,
	citecolor=green,
	linkcolor=blue,
}
\usepackage{color,hyperref}
\definecolor{mediumorchid}{rgb}{0.73, 0.33, 0.83}
\definecolor{twilightlavender}{rgb}{0.54, 0.29, 0.42}
\definecolor{purplemountainmajesty}{rgb}{0.59, 0.47, 0.71}
\definecolor{brightmaroon}{rgb}{0.76, 0.13, 0.28}
\definecolor{richmaroon}{rgb}{0.69, 0.19, 0.38}
\definecolor{midnightblue}{rgb}{0.1, 0.1, 0.44}
\definecolor{rosebonbon}{rgb}{0.98, 0.26, 0.62}

\title{Counting $\mathcal{N}=8$ Black Holes as Algebraic Varieties}

\author{Abhishek Chowdhury,}
\author{Sourav Maji} 
\affiliation{ School of Basic Sciences,\\ Indian Institute of Technology Bhubaneswar, \\Jatni, Khurda, Odisha, 752050, India}
\emailAdd{achowdhury@iitbbs.ac.in}
\emailAdd{sm89@iitbbs.ac.in}

 \abstract{We calculate the helicity trace index $B_{14}$ for $\mathcal{N}=8$ pure D--brane black holes using various techniques of computational algebraic geometry and find perfect agreement with the existing results in the literature. For these black holes, microstate counting is equivalent to finding the number of supersymmetric vacua of a multi--variable supersymmetric quantum mechanics which in turn is equivalent to solving a set of multi--variable polynomial equations modulo gauge symmetries. 
 We explore four different techniques to solve a set of polynomial equations, namely Newton Polytopes, Homotopy continuation, Monodromy and Hilbert series. The first three methods rely on a  mixture of symbolic and high precision numerics whereas the Hilbert series is symbolic and admit a gauge invariant analysis. Furthermore, exploiting various exchange symmetries, we show that quartic and higher order terms are absent in the potential, which if present would have spoiled the counting. Incorporating recent developments in algebraic geometry focusing on computational algorithms, we have extended the scope of one of the authors previous works \cite{Chowdhury:2014yca,Chowdhury:2015gbk} and presented a new perspective for the black hole microstate counting problem. This further establishes the pure D--brane system as a consistent model, bringing us a step closer to $\mathcal{N} =2$ black hole microstate counting.}

\arxivnumber{2311.04786}

\begin{document}

\maketitle

\section{Introduction}  \label{s1}

The interplay between mathematics and physics has been very rewarding both from the perspective of mathematicians and physicists alike. Substantial contribution can be attributed to String theory which represents a framework to quantize gravity within the realm of theoretical physics and enjoy deep connections with mathematics, especially in the areas of topology \cite{Vonk2005AMO}, algebraic geometry \cite{Hori:2003ic} and number theory \cite{hoker2022LecturesOM}. String theory in turn, has advanced well owing to the extensive studies on black holes as theoretical laboratories suitable for controlled theoretical predictions. An important milestone in string theory is the validation of its internal mathematical consistency spearheaded by the precise resolution of Bekenstein--Hawking entropy \cite{Hawking:1975vcx,Bekenstein1973BlackHA} as degeneracy of BPS states representing the internal degrees of freedom of a black hole. Over the past two decades, significant progress has been made towards counting BPS states of supersymmetric extremal black holes in $\mathcal{N}=8$ \cite{n8Sen:2008ta,n8Sen:2009gy,n8Shih:2005qf,Chowdhury:2014yca,Chowdhury:2015gbk,Pioline:2005vi}, $\mathcal{N}=4$ \cite{n4David:2006ji,n4David:2006ru,n4Dijkgraaf:1996it,n4Gaiotto:2005hc,n4Jatkar:2005bh,n4Shih:2005uc,Sen:entropyfunction,Mandal:2010cj} and $\mathcal{N}=2$ \cite{n2Manschot:2011xc,n2deBoer:2008zn} theories. \\
For a typical four dimensional black hole, a weak coupling microscopic description would have a higher dimensional type II or Heterotic string theory appropriately compactified on a tori, $K3$ or \textit{Calabi--Yau} spaces (preserving 8 or more supercharges) with a collection of intersecting D--branes and strings wrapping around 1--cycles forming a bound state carrying charges (in the process breaking most of the SUSY). The SUSY preserving BPS state degeneracy ($d$) of its low energy dynamics is captured by computing a supersymmetric/Witten index of the low energy excitations. On the macroscopic side, where gravity is relatively strong, the bound state is interpreted as a black hole solution of an appropriate supergravity theory with entropy ($S_{BH}$) proportional to the area of its event horizon. Barring the important caveat of degeneracy vs index \cite{Mandal:2010cj}, the invariance of the index under deformations results in $d \approx e^{S_{BH}}$. \\
Near horizon analysis of BPS black holes carrying four unbroken SUSY suggests that the microstates of single center black holes carry zero angular momentum leading to the positivity of the index \cite{Sen:2009vz, Dabholkar:2010rm} \footnote{The microstates are tensor product of a BPS multiplet obtained by quantizing goldstino fermion zero modes and a singlet representation of the $SU(2)$ rotation group.}. It is however difficult to test this conjecture directly as the regimes of microstate counting and black hole description are different but positivity of the index has been checked in many examples \cite{Sen:2011ktd, Bringmann:2012zr, Govindarajan2022PositivityOD, Chattopadhyaya2021HorizonSA} and no counterexamples has been found. In previous works of one of the authors \cite{Chowdhury:2015gbk} a `zero angular momentum conjecture' was put forward which states that at a generic point in the moduli space in both macroscopic and microscopic regimes, the spectrum of BPS states will carry zero angular momentum \footnote{In a non--generic or special point in the moduli space where for example most moduli would take zero values, the spectrum of BPS states have different angular momenta and an unequal cancellation of the positive and negative contribution to the index gives a net positive contribution.}. To test this conjecture it is best to start with $\mathcal{N}=8$ theories where it is known that only single centers contribute to the index \cite{n8Sen:2008ta} and we don't have to the deal with the complications arising from wall crossing and two or multi--center contributions to the index as are the case for $\mathcal{N}=4$ \cite{Sen:2007vb, Dabholkar:2007vk, Sen:2007pg, Chowdhury:2019mnb, LopesCardoso:2020pmp, Chowdhury:2012jq} and $\mathcal{N}=2$ \cite{Denef:2007vg, n2Manschot:2011xc} theories. \\
In this paper we shall focus on a pure D--brane D2--D2--D2--D6 system in type IIA theory compactified on $T^6$ that is dual to a D1--D5--P--KK monopole system in type IIB on $T^6$ for which the microscopic index is known \cite{n8Shih:2005qf}. The original motivation to study this system as a gateway to the counting of black holes in $\mathcal{N}=2$ string theories compactified on Calabi--Yau manifolds still stands \cite{Shmakova1996CalabiYauBH, Gaiotto2005NewCB, Maldacena:1997de}. As an opening move, we extend the analysis of \cite{Chowdhury:2014yca,Chowdhury:2015gbk} where it was shown that for abelian charge $(1,1,1,1)$ and non--abelian charges $(1,1,1,2)$ \& $(1,1,1,3)$ the SUSY vacua manifold for the $0+1$ dimensional quantum mechanics with support on the intersection of the four branes is a collection of 12, 56 and 208 points respectively. These results in support of  the proposed conjecture were found by solving the F--terms equations which are multi--variable polynomials using Mathematica \cite{Mathematica} which we now know to be using an numerical algorithm based on \textit{Gr$\ddot{o}$bner basis}. Using myriad sophisticated techniques developed in computational algebraic geometry to deal with \textit{algebraic} or \textit{projective varieties} \cite{Bliss2018MonodromySS, Sommese2005TheNS, jejjala:2006jb} we can extend these results to other non--abelian charges $(1,1,1,N)$ but we will focus on $(1,1,1,4)$ which gives a SUSY vacua of 684 points again in support of the conjecture. For the known results, some of these methods are more than 100 times faster than Mathematica, even on a single core \footnote{We have a Dell Workstation with two Intel Xeon Gold 6258R processors (56 cores) and 128 GB of RAM. The main bottleneck in the algorithmic implementation of most of these techniques is the RAM.}. We have put emphasis on the Hilbert series methods as they are symbolically exact and allows us to extract the solutions count without explicit knowledge about the actual solutions. Along the way, we have provided an explicit check for the vacua count by computing the \textit{Hessian} around the critical points of the potential and argued the consequences of introducing higher order terms in the potential.\\  
A companion goal of this paper is to highlight the interplay between a robust, sufficiently complex physical model with various symmetries and techniques of  computational algebraic geometry. The application of algebraic geometry to black holes in string theory is largely due to the study of Calabi--Yau \cite{cyHe:2017gam,Aspinwall1994StringTO}, structure of various moduli spaces \cite{Mehta2012NumericalAG,Mehta2011NumericalPH} and mirror symmetry \cite{Hori:2003ic} but to the best of our knowledge, application of numerical and symbolically exact algorithms developed in algebraic geometry have not featured in any previous works of microstate counting in black holes. Among the many techniques we studied \cite{Sommese2005TheNS,allgower2003introduction,Bliss2018MonodromySS}, in this paper we present four of them, namely the Newton polytope, Homotopy continuation, Monodromy and the Hilbert series based on the suitability to our problem, parallelizability and wider applicability. These methods can also be applied to a large class of cost function/energy minimization problems where the landscape of low lying critical points/states is expected to grow exponentially with respect to some parameters, like the string vacua landscape \cite{vlHe:2017set,Mehta2012NumericalAG} and the cost functions in machine learning \cite{costWang2020ACS}. For mathematicians, these D--brane models presents an infinite class of 
 parametric weighted projective varieties whose number of critical points/intersection number are coefficients of a known generating function (a weak Jacobi form)  and for large values of the parameters $N_1$, $N_2$, $N_3$ and $N_4$ (charges of the black hole), the count is $ \displaystyle d\approx e^{2\pi\sqrt{N_1 N_2 N_3 N_4}-2 \ln (4 N_1 N_2 N_3 N_4)}$.  \\
The rest of the paper is organized as follows. In section \ref{review} we review the D--brane model focusing mainly on the potentials, the symmetries and the nature of vacua manifold. In section \ref{algbericbh} we briefly review the mathematical ideas behind the computational algebraic geometry techniques of Newton polytope, Homotopy continuation, Monodromy and Hilbert series and then discuss our strategies to apply them in the context of solving the F--term equations arising in our models, along with the results. In section \ref{hilbertbh} we continue with Hilbert series, providing more details on gauge (un-)fixing and discuss the results. In section \ref{quarticz} we show that addition of a quartic term to our potential will potentially destroy the U--duality. We give symmetry arguments to rule out such terms from the superpotential. We conclude in section \ref{discussion} with some implications of the zero angular momentum conjecture along with a broad overview of the applicability and interplay of these methods with various fields of study. There are three appendices, in appendix \ref{detailedhilbert1111} we discuss in detail the contour integrals involved in the Hilbert series for charge $(1,1,1,1)$, in appendix \ref{14helicitytrace} we summarize the extraction of the microstate counts from its generating function, the weak Jacobi form and tabulate the relevent results, in appendix \ref{gaugecharges} we tabulate the charge tables used in the Hilbert series computations for various abelian and non--abelian charge configurations. The paper ends with the references.

\section{Review of the D--brane System} \label{review}
In this section we will review the D--brane model proposed in \cite{Chowdhury:2014yca,Chowdhury:2015gbk} and give a brief summary of its action, the potentials, various symmetries and finally our expectation from the supersymmetric vacua manifold. The pure D--brane D2--D2-D2--D6 system is a D--brane intersection model in type IIA string theory compactified on $T^6$ with $N_1$ D2--brane wrapping the  $(x^4, \, x^5)$ directions, $N_2$ D2--brane along $(x^6, \, x^7)$ directions, $N_3$ D2--brane along $(x^8, \, x^9)$ directions and finally $N_4$ D6--brane along $(x^4,\cdots,x^9)$ directions. As the system breaks 28 out of 32 supersymmetries, it can be described in terms of $\mathcal{N}=1$ supersymmetry in $3+1$ dimensions and we can use the $\mathcal{N}=1$ superfield formalism to organize the field content as $\mathcal{N}=1$ multiplets. \\
Associated with each D--brane is a low energy effective theory which in this case is the dimensional reduction from $3+1$ to $0+1$ 
 dimensions of a $\mathcal{N}=4$ supersymmetric $U(N)$ Yang--Mills theory coming from string fluctuations starting and ending on the same stack of branes. It has one $\mathcal{N}=1$ vector multiplet $V^{(k)}$ and three $\mathcal{N}=1$ chiral multiplets $\Phi_1^{(k)}$, $\Phi_2^{(k)}$ and $\Phi_3^{(k)}$, where $k$ labels the brane index. On the other hand, the strings between a pair of branes, starting at one brane and ending on another give rise to a $\mathcal{N}=2$ hyper--multiplet, or equivalently two $\mathcal{N}=1$ chiral multiplets $Z^{(k\ell)}$ and $Z^{(\ell k)}$.

\subsection{Action and Potentials}\label{potentials}

Assuming the six circles of $T^6$ are orthonormal to each other with radii $\sqrt{\alpha'}$, which we will set to $\alpha'=1$, the action of the pure D--brane system is given by,
\begin{equation}\label{lagrangian}
S_{\text {kinetic }}+\int d x^0\left[\int d^4 \theta \sum_{k=1}^4 \sum_{\substack{\ell=1 \\ \ell \neq k}}^4\left\{\bar{Z}^{(k \ell)} e^{2 V^{(\ell)}-2 V^{(k)}} Z^{(k \ell)}\right\}+\int d^2 \theta \mathcal{W}+\int d^2 \bar{\theta} \overline{\mathcal{W}}\right] \,,
\end{equation}
where $S_{\text {kinetic }}$ is the standard kinetic term involving the chiral and vector superfields. The interaction superpotential $\mathcal{W}$ has three different terms, $\mathcal{W}_1$, $\mathcal{W}_2$ and $\mathcal{W}_3$. $\mathcal{W}_1$ describes the coupling between the superfields $Z^{(k\ell)}$ and $\Phi^{(k)}_{m}$ and it is of the following form,
\begin{equation}
\begin{aligned}
\mathcal{W}_1=\sqrt{2} & {\left[\sum_{k, \ell, m=1}^3 \varepsilon^{k \ell m} \operatorname{Tr}\left(\Phi_m^{(k)} Z^{(k \ell)} Z^{(\ell k)}\right)+\sum_{k=1}^3 \operatorname{Tr}\left(Z^{(4 k)} \Phi_k^{(k)} Z^{(k 4)}\right)\right.}\\
& \left.-\sum_{k=1}^3 \operatorname{Tr}\left(\Phi_k^{(4)} Z^{(4 k)} Z^{(k 4)}\right)\right] \,,
\end{aligned}
\end{equation}
where $\varepsilon^{k \ell m}$ is the Levi--Civita symbol with $\varepsilon^{123}=1$. $\mathcal{W}_2$ describes the cubic self--coupling between the $Z^{(k \ell)}$'s and takes the form,
\begin{equation}
\label{eq:cubic}
\mathcal{W}_2=\sqrt{2} \, C \left[\sum_{\substack{k, l, m=1 \\ k<\ell, m ; l \neq m}}^4 (-1)^{{\delta_{k1}} {\delta_{\ell 3}} {\delta_{m 4}}}\operatorname{Tr}\left(Z^{(k \ell)} Z^{(\ell m)} Z^{(m k)}\right)\right]\,.
\end{equation}
If small background values of the off--diagonal components of the metric and 2--form $B$--fields are switched on, we get an extra superpotential term,
\begin{equation}\label{eq:w3}
\mathcal{W}_3=\sqrt{2}\left[\sum_{k, \ell, m=1}^3 c^{(k \ell)} \varepsilon^{k \ell m} N_{\ell} \operatorname{Tr}\left(\Phi_m^{(k)}\right)+\sum_{k=1}^3 c^{(k 4)}\left[N_4 \operatorname{Tr}\left(\Phi_k^{(k)}\right)-N_k \operatorname{Tr}\left(\Phi_k^{(4)}\right)\right]\right]
\end{equation}
There is also an additional superpotential
\begin{equation}\label{SYM}
\mathcal{W}_4=-\sqrt{2} \sum_{k=1}^4 (-1)^{\delta_{k2}}\operatorname{Tr}\left(\Phi_1^{(k)}\left[\Phi_2^{(k)}, \Phi_3^{(k)}\right]\right)\,.
\end{equation}
 Here, $C=1$, is a constant whose value can be computed \cite{Chowdhury:2015gbk} by analyzing the coupling between open strings stretched between different branes and $c^{(k)}$ are Fayet--Iliopoulos (FI) parameters satisfying $ \displaystyle \sum_{k=1}^4 c^{(k)}N_{k}=0$. By comparing the mass of open strings stretched between a pair of branes and the renormalized mass determined from the quadratic term involving the fields $Z^{(k\ell)}$ and $Z^{(\ell k)}$, we can write $c^{(k)}$ and $c^{(k \ell)}$ in terms of the off--diagonal components of metric $g_{mn}$ and 2--form field $b_{mn}$. The explicit expressions are given in the appendix A of \cite{Chowdhury:2014yca}.\\
Writing the action in terms of the component field, we get three different potential terms,
\begin{align}
V_{\text {gauge }}= & \sum_{k=1}^4 \sum_{\substack{k=1 \\
l \neq k}}^4 \sum_{i=1}^3 \operatorname{Tr}\left[\left(X_i^{(k)} Z^{(k \ell)}-Z^{(k \ell)} X_i^{(\ell)}\right)^{\dagger}\left(X_i^{(k)} Z^{(k \ell)}-Z^{(k \ell)} X_i^{(\ell)}\right)\right] \\
& +\sum_{k=1}^4 \sum_{i, j=1}^3 \operatorname{Tr}\left(\left[X_i^{(k)}, \Phi_j^{(k)}\right]^{\dagger}\left[X_i^{(k)}, \Phi_j^{(k)}\right]\right) \notag \\
& +\frac{1}{4} \sum_{k=1}^4 \sum_{i, j=1}^3 \operatorname{Tr}\left(\left[X_i^{(k)}, X_j^{(k)}\right]^{\dagger}\left[X_i^{(k)}, X_j^{(k)}\right]\right) \notag \, .
\end{align}
We abuse the notation here as the scalar components are being denoted with the same name $Z^{(kl)}$, $\Phi_i^{(k)}$ etc. as superfields. $X_i^{(k)}$ are gauge fields arising from the dimension reduction of the vector fields of the vector multiplets, and we choose the gauge $A_{0}=0$. The D--term potential takes the form,
\begin{equation} \label{dtermpotential}
V_D=\frac{1}{2} \sum_{k=1}^4 \operatorname{Tr}\left[\left(\sum_{\substack{l=1 \\ l \neq k}}^4 Z^{(k \ell)} Z^{(k \ell) \dagger}-\sum_{\substack{i=1 \\ i \neq k}}^4 Z^{(\ell k) \dagger} Z^{(\ell k)}+\sum_{i=1}^3\left[\Phi_i^{(k)}, \Phi_i^{(k) \dagger}\right]-c^{(k)} I_{N_k}\right)^2\right]\,
\end{equation}
and the F--term potential is,
\begin{equation}
V_F=\sum_{k=1}^4 \sum_{i=1}^3\left|\frac{\partial W}{\partial \Phi_i^{(k)}}\right|^2+\sum_{k=1}^4 \sum_{\substack{\ell=1 \\ \ell \neq k}}^4\left|\frac{\partial W}{\partial Z^{(k \ell)}}\right|^2 \,.
\end{equation}

\subsection{Symmetries} \label{symmetries}
The above Lagrangian has various continuous and discrete symmetries which we list here. Throughout the rest of the paper we have used the Gauge symmetries and the shift symmetries to constraint the system before ateempting to solve for the SUSY vacua. In section 
 \ref{quarticz} we used the discrete symmetries to show that quartic and higher order terms are not allowed in the superpotential.
\subsubsection*{Gauge symmetries}
The system originally has  a $U(N_1) \times U(N_2) \times U(N_3) \times U(N_4)$ gauge symmetry. As we have partially gauge fixed to $A_0=0$, we only focus on the global part of the gauge symmetry, with various fields transforming as $Z^{(k\ell)} \rightarrow U(N_k)\, Z^{(k \ell)} \,U^{-1}(N_\ell)$ and $\Phi^{(k)}_i \rightarrow U(N_k)\, \Phi^{(k)}_i \, U^{-1}(N_k)$. In the later sections we will discuss various ways to implement gauge fixing for various charge configurations. 

\subsubsection*{Shift symmetries}
Type IIA string theory in ten space--time dimensions has 32 supercharges. The Lagrangian described in the previous section is essentially that of a SUSY quantum mechanics which has support on the intersection of four D--branes. It breaks 28 out the 32 supercharges, resulting in 28 Goldstinos. The remaining four supercharges ensure that the Goldstinos have their bosonic partners, pairing up as $\mathcal{N}=1$ massless hyper--multiplets. These bosonic zero modes are the consequence of spontaneous symmetry breaking due the insertion of D--branes. We have called them \textit{shift symmetries} and are given by,
\begin{equation} \label{eflatgen}
\begin{aligned}
& \Phi^{(k)}_m \to \Phi^{(k)}_m+\xi_m I_{N_k},  \quad \hbox{for} 
\quad 1\le k \le 3, \quad k \ne m; \quad  1\le m\le 3 ,\, \\
& \Phi^{(k)}_k \to \Phi^{(k)}_k + \zeta_k I_{N_k}, \quad \Phi^{(4)}_k \to
\Phi^{(4)}_k+\zeta_k I_{N_4}, \quad \hbox{for} \quad 1\le k\le 3 ,\, \\
& X^{(k)}_i \to X^{(k)}_i + a_i \, I_{N_k}\, , \quad \hbox{for} \quad 1\le i\le 3\, .
\end{aligned}
\end{equation}
\noindent where $\xi_m$ and $\zeta_k$ are arbitrary complex numbers and $a_i$ are arbitrary real numbers representing overall translation of the system along the non--compact directions. In the black hole microstate counting programme, we mod out by these hyper--multiplets while computing the $B_{14}$ Helicity trace index. Therefore, we have to `gauge fix' these flat directions of the potential to compute the SUSY ground states \footnote{As a consequence of modding out by translations in the non--compact directions, the gauge symmetries have to be modded out 
 by an overall diagonal $U(1)$ i.e. $(U(N_1) \times U(N_2) \times U(N_3) \times U(N_4))/U(1). $ }. Throughout the paper while doing computations, we have chosen 
 \begin{equation}\label{eq:phase}
     \Phi^{1}_{1}=0, \, \Phi^{1}_{2}=0, \, \Phi^{1}_{3}=0, \, \Phi^{2}_{1}=0, \, \Phi^{2}_{2}=0 \, \text{and} \,  \Phi^{3}_{3}=0 \, .
 \end{equation}

\subsubsection*{Discrete symmetries}
Type IIA theory on $T^6$ exhibits  various discrete  \textit{exchange symmetries}. These symmetries constrain the form of the superpotentials $\mathcal{W}_2$ \eqref{eq:cubic} and $\mathcal{W}_4$ \eqref{SYM} . Later, in section \ref{quarticz}, we will use these symmetries to show that the superpotential doesn't admit additional terms of quartic order and beyond. 

\begin{itemize}
    \item \textbf{First Exchange Symmetry}: We exchange the D2--brane stacks 1 and 2 i.e. $x^4 \leftrightarrow x^6$, $x^5 \leftrightarrow x^7$ and $N_1 \leftrightarrow$ $N_2$. Moreover, $c^{(34)}$ changes sign, $c^{(12)}$ remains unchanged, and $c^{(1 i)}$ and $c^{(2 i)}$ get exchanged for $i=3,4$. The superpotentials transform as  $\mathcal{W}_i \rightarrow-\mathcal{W}_i$ for $1 \leq i \leq 4 $ and various field transform in the following way, 
\begin{equation}
\begin{aligned}
& \left(\Phi_3^{(4)}, \Phi_3^{(3)}\right) \rightarrow\left(\Phi_3^{(4)}, \Phi_3^{(3)}\right), \quad\left(\Phi_3^{(2)}, \Phi_3^{(1)}\right) \rightarrow\left(\Phi_3^{(1)}, \Phi_3^{(2)}\right), \\
& \left(\Phi_1^{(2)}, \Phi_1^{(3)}\right) \leftrightarrow\left(\Phi_2^{(1)}, \Phi_2^{(3)}\right), \quad\left(\Phi_1^{(4)}, \Phi_1^{(1)}\right) \leftrightarrow-\left(\Phi_2^{(4)}, \Phi_2^{(2)}\right), \\
& Z^{(34)} \rightarrow Z^{(34)}, \quad Z^{(i 1)} \leftrightarrow-Z^{(i 2)}, \quad Z^{(1 i)} \leftrightarrow-Z^{(2 i)}, \quad \text { for } i=3,4, \\
& Z^{(12)} \leftrightarrow-Z^{(21)}, \quad Z^{(43)} \rightarrow-Z^{(43)} \, .
\end{aligned}
\end{equation}
     \item \textbf{Second Exchange Symmetry}: We exchange the D2--brane stacks 2 and 3 i.e. $x^6 \leftrightarrow x^8$, $x^7 \leftrightarrow x^9$ and $N_2 \leftrightarrow N_3$. Moreover, we have $c^{(14)} \rightarrow-c^{(14)}, c^{(23)} \rightarrow c^{(23)}, c^{(12)} \leftrightarrow c^{(13)}$ and $c^{(24)} \leftrightarrow c^{(34)}$. The superpotentials transform as $\mathcal{W}_i \rightarrow-\mathcal{W}_i$ for $1 \leq i \leq 4 $ and various fields transform in the following way, 
\begin{equation}
\begin{aligned}
& \left(\Phi_1^{(4)}, \Phi_1^{(1)}\right) \rightarrow\left(\Phi_1^{(4)}, \Phi_1^{(1)}\right), \quad\left(\Phi_3^{(1)}, \Phi_3^{(2)}\right) \leftrightarrow\left(\Phi_2^{(1)}, \Phi_2^{(3)}\right), \\
& \left(\Phi_1^{(2)}, \Phi_1^{(3)}\right) \rightarrow\left(\Phi_1^{(3)}, \Phi_1^{(2)}\right), \quad\left(\Phi_2^{(4)}, \Phi_2^{(2)}\right) \leftrightarrow-\left(\Phi_3^{(4)}, \Phi_3^{(3)}\right), \\
& Z^{(41)} \rightarrow Z^{(41)}, \quad Z^{(2 i)} \leftrightarrow-Z^{(3 i)}, \quad Z^{(i 2)} \leftrightarrow-Z^{(i 3)}, \quad \text { for } i=1,4, \\
& Z^{(32)} \leftrightarrow-Z^{(23)}, \quad Z^{(14)} \rightarrow-Z^{(14)} \,.
\end{aligned}
\end{equation}   
     \item \textbf{Third Exchange Symmetry}:  We exchange the D2--brane stacks 1 and 4. Here, it is best to perform a T--duality transformation along 6--7--8--9 directions, exchange $x^6 \leftrightarrow x^8$, $x^7 \leftrightarrow x^9$ and at the same time exchange $N_1$ and $N_4$. Moreover, $c^{(14)}$ remains unchanged, $c^{(23)}$ changes sign, $c^{(12)} \leftrightarrow i c^{(24)}$ and $c^{(13)} \leftrightarrow i c^{(34)}$. The superpotentials transform as $\mathcal{W}_i$ to $-\mathcal{W}_i$ for $1 \leq i \leq 4$ and various fields transforms as follows:
\begin{equation}
\begin{aligned}
& \left(\Phi_1^{(4)}, \Phi_1^{(1)}\right) \rightarrow\left(\Phi_1^{(1)}, \Phi_1^{(4)}\right), \quad\left(\Phi_2^{(1)}, \Phi_2^{(3)}\right) \leftrightarrow i\left(\Phi_3^{(4)}, \Phi_3^{(3)}\right), \\
& \left(\Phi_1^{(2)}, \Phi_1^{(3)}\right) \rightarrow\left(\Phi_1^{(2)}, \Phi_1^{(3)}\right), \quad\left(\Phi_2^{(4)}, \Phi_2^{(2)}\right) \leftrightarrow i\left(\Phi_3^{(1)}, \Phi_3^{(2)}\right), \\
& Z^{(12)} \leftrightarrow-Z^{(42)}, \quad Z^{(13)} \leftrightarrow-i Z^{(43)}, \quad Z^{(21)} \leftrightarrow-i Z^{(24)}, \quad Z^{(31)} \leftrightarrow-Z^{(34)}, \\
& Z^{(14)} \leftrightarrow i Z^{(41)}, \quad Z^{(32)} \rightarrow Z^{(32)}, \quad Z^{(23)} \rightarrow-Z^{(23)}\, .
\end{aligned}
\end{equation}    
\end{itemize}
Compositions of the above three symmetries generate all other exchange symmetries. There are other symmetries, for example, the worldsheet parity symmetry, under which the NS--NS 2--form field changes sign. This is however not a symmetry of type IIA string theory unless it is accompanied by the parity transformation along the non--compact directions $(- 1)^{F_L}$, then this becomes a symmetry of the theory. The details can be found in appendix A of \cite{Chowdhury:2015gbk}.

\subsection{Supersymmetric Vacua} \label{susyvacua}
The Lagrangian as described in section \ref{potentials} has a potential, $V=V_{\text{gauge}}+V_{D}+V_{F}$. The vacua manifold comprises of supersymmetric solutions which are stationary solutions with energy $E=V=0$. As the potential $V$ is a sum of square terms, vanishing of the potential implies that each individual term equals to zero. Our first agenda is to reduce as much as possible the number of fields over which we have to minimize the potential. To that end, we pick the $V_{gauge}$ term, diagonalize the gauge fields $X^{(k)}_i$ using the gauge symmetry,  $U(N_1) \times U(N_2) \times U(N_3) \times U(N_4)$ and following the arguments of appendix A of \cite{Chowdhury:2015gbk}, set all $X^{(k)}_i=0$ for $V_{gauge}=0$. We are now left with the $V_{D}$ and $V_{F}$ terms in the potentials with active fields $Z^{(k\ell)}$ and $\Phi^{(k)}_i$. Upon complexifying the original gauge invariance to $ (GL(N_1,\mathcal{C}^{*}) \times  GL(N_2,\mathcal{C}^{*}) \times  GL(N_3,\mathcal{C}^{*}) \times GL(N_4,\mathcal{C}^{*}))/ GL(1,\mathcal{C}^{*})$, we can effectively set $V_{D}$ to zero \footnote{For example, in the abelian case, the $Z^{(k \ell)}$ span a \textit{projective variety} but as no coordinates can go to zero as long as $c^{(k \ell)} \neq 0\,$, it is a \textit{toric variety}. In case coordinates are allowed to go to zero, we have to perform the counting in different coordinate patches and then take a union of the results.}. We can now focus only on the vanishing of the F--term potential, which results in the following F--term equations to be solved to obtain the vacua manifold.
\begin{equation} \label{eq:fterm1}
\begin{aligned}
Z^{(k \ell)} Z^{(\ell k)} & =-c^{(k \ell)} N_{\ell} I_{N_k}+\left[\Phi_k^{(k)}, \Phi_{\ell}^{(k)}\right] \quad \text { for } \quad 1 \leq k, \ell \leq 3, \\
Z^{(k 4)} Z^{(4 k)} & =-c^{(k 4)} N_4 I_{N_k}+\sum_{\ell, m=1}^3 \varepsilon^{k \ell m} \Phi_{\ell}^{(k)} \Phi_m^{(k)}, \quad 1 \leq k \leq 3, \\
Z^{(4 k)} Z^{(k 4)} & =-c^{(k 4)} N_k I_{N_4}-\sum_{\ell, m=1}^3 \varepsilon^{k \ell m} \Phi_{\ell}^{(4)} \Phi_m^{(4)}, \quad 1 \leq k \leq 3,
\end{aligned}
\end{equation}
\begin{align}
\label{eq:fterm2}
& \sum_{m=1}^3 \varepsilon^{k \ell m}\left(Z^{(\ell k)} \Phi_m^{(k)}-\Phi_m^{(\ell)} Z^{(\ell k)}\right)+C \sum_{\substack{m=1  \\
m \neq k, \ell}}^4 Z^{(\ell m)} Z^{(m k)} (-1)^{\delta_{k 1}\delta_{\ell 3}\delta_{m 4} }=0 \text { for } 1 \leq k, \ell \leq 3, \notag \\
& \left(\Phi_k^{(k)} Z^{(k 4)}-Z^{(k 4)} \Phi_k^{(4)}\right)+C \sum_{\substack{\ell=1 \\
\ell \neq k}}^3 Z^{(k \ell)} Z^{(\ell 4)} (-1)^{\delta_{k 1}\delta_{\ell 3}}=0 \quad \text { for } \quad 1 \leq k \leq 3,  \\
& \left(Z^{(4 k)} \Phi_k^{(k)}-\Phi_k^{(4)} Z^{(4 k)}\right)+C \sum_{\substack{m=1 \\
m \neq k}}^3 Z^{(4 m)} Z^{(m k)} (-1)^{\delta_{m 1}\delta_{k 3}}=0 \quad \text {for} \quad 1 \leq k \leq 3 \, . \notag
\end{align}
To better understand the expectations from the vacua manifold we look at a standard \textit{Landau Ginsberg} theory with $n$ chiral hyper--multiplets and four supersymmetries \cite{158d02b66ce6495492b1b30126556c95, Witten:1993yc}. The Lagrangian is given by
\begin{equation}\label{landauginzberg}
\mathcal{L}=\mathcal{L}_{\text {kinetic }}  -\frac{1}{4}\sum_{i=1}^n\left|\partial_i \mathcal{W}\right|^2 
-\frac{1}{2} \sum_{i, j}\left(\partial_i \partial_j \mathcal{W} \Psi^i \tilde{\Psi}^j+\bar{\partial}_{i} \bar{\partial}_{j} \bar{\mathcal{W}} \overline{\tilde{\Psi}}^{i} \bar{\Psi}^{j}\right),
\end{equation}
where the superpotential $\mathcal{W}(z)=-(h(x, y)+ i f(x, y))$, is a holomorphic function of the complex fields, $z_i=x_i+i y_i\,$. Our lagrangian, reduces to the above lagrangian after we remove the shift symmetries, put the the gauge fields $X_i^{(k)} =0$ and complexify the gauge groups to set $V_D=0$. Albeit, we still have to mod out by the complexified gauge orbits. The critical points of the superpotential are associated with the semi--classical ground states of the theory, i.e.
\begin{equation}
\partial_i \mathcal{W}=0 \text { for all } i=1, \ldots, n
\end{equation}
which are critical points of $h(x,y)$ too. For suitably generic $h(x,y)$ there are no degenerate critical points and the Witten Index is given by 
\begin{equation}
\operatorname{Tr}(-1)^F e^{-\beta H}=\sum_{(X,Y)}(-1)^{\mu(X,Y)}\, ,
\end{equation}
where $(X,Y)$ are the critical points of $h(x,y)$. Here, $\mu (X,Y)$, is the \textit{Morse Index}, 
\begin{equation}
\mu(X,Y)=\text {\# of --ve eigenvalues of the \textit{Hessian} }=-\partial_i \partial_j h(X,Y) \, .
\end{equation}
The ground state wave--function attached to the critical point $ (X,Y)$ has $p = \mu (X,Y)$ excited
fermions, i.e. it is a $p$--form \footnote{Generically, some semi--classical ground states around critical points of $h(X,Y)$ can get lifted to non--zero energy and don't correspond to true $E = 0$ states. But, as they get lifted in boson--fermion pairs the Witten index remains the same.}. \\
Fortunately, for the SUSY Landau Ginsberg models $h(x,y)$ is not completely generic. For a critical point at say $(X,Y)=(0,0)$, after suitable diagonalization 
\begin{equation}
\mathcal{W}(z) \approx \sum_i w_i\left(z^i\right)^2+\ldots
\end{equation}
and in terms of $h(x, y)$ it is
\begin{equation}
h(x, y)=-\operatorname{Re} \mathcal{W}(z) \approx -\sum_i w_i\left(\left(x^i\right)^2-\left(y^i\right)^2\right)+\ldots
\end{equation}
Holomorphy of $\mathcal{W}$ ensures that for every +ve eigenvalue there is a paired --ve eigenvalue of the Hessian of $h$, which in turn  ensures that every critical point has Morse index $n$ contributing exactly the same to the Witten index, 
\begin{equation}
\operatorname{Tr}(-1)^F e^{-\beta H}=(-1)^n \times \text {\# of critical points $(X,Y)$ of} \,\,\mathcal{W}
\end{equation}
i.e. all critical point of $\mathcal{W}$ are 
true $E = 0$ ground states of the quantum theory. No semi--classical ground state gets lifted by quantum tunneling/instanton effects, hence in our case degeneracy of critical points equals Witten index up to a sign \footnote{Following the arguments involving the \textit{Lefschetz} $SU(2)$ and \textit{middle cohomology} of  the vacua manifold, in section 2 of \cite{Chowdhury:2015gbk}, for the cases discussed in this paper the vacua manifolds are indeed a collection of isolated points and hence the corresponding BPS states carry zero angular momentum.}.\\
In the next section we will solve the set of F--term equations by treating them as polynomial varieties and discuss various numerical and exact methods to solve a set of equations. We have also computed the Hessian for various D--brane charges and explicitly verified the above claim.

\section{Counting Black Holes as Algebraic Varieties} \label{algbericbh}
In this section we discuss various ways to solve the F--term and D--term equations to get information about the vacuum moduli space of the the pure D--brane system. The vacuum moduli space is best described as an algebraic variety. The geometries we are interested in  are \textit{affine varieties}, which are points, curves, surfaces and higher dimensional objects defined by polynomial equations over the complex field $\mathbb{C}$ \footnote{For computer algorithms, we mostly used rational number field $\mathbb{Q}$. Sometimes, using a Galois field of integers mod $p$ where $p$ is a prime number speeds up the computations.}. The set of all polynomials over the field $\mathbb{C}$ form a commutative ring $\mathbb{C}[x_1, \ldots x_n]$, referred to as a \textit{polynomial ring} \footnote{$\mathbb{C}[x_1,\ldots x_n]$ satisfies all field axioms except for the existence of multiplicative inverses as $1/x$ is not a polynomial.}. Given the field $\mathbb{C}$ and $n \in \mathbb{Z}_{\geq 0}$ , we can define the $n$--dimensional \textit{affine space} over $\mathbb{C}$ to be the set 
\begin{equation}
    \mathbb{C}^n = \left \{(a_1, \ldots, a_n) \,|\, a_1, \cdots, a_n \in \mathbb{C} \right \}\,.
\end{equation}
The link between algebra and geometry comes from the fact that a polynomial $f(x_1,\ldots, x_n)$  can be regarded as a function over the affine space. The set of all simultaneous solutions $\left(a_1,\ldots, a_n\right) \in \mathbb{C}^n$ of a system of polynomial equations
\begin{equation}\label{poly}
\begin{aligned}
f_1\left(x_1, \ldots, x_n\right) & =0 \\
f_2\left(x_1, \ldots, x_n\right) & =0 \\
& \vdots \\
f_s\left(x_1, \ldots, x_n\right) & =0
\end{aligned}
\end{equation}
is an \textit{affine variety}, $V= \mathcal{V}\left(f_1, \ldots, f_s\right)$. To make the connection to our system, we consider the fields $Z^{(k\ell)}$ and $\Phi^{(k)}_i$ as the coordinates $x_1, \ldots, x_n$ and the F--term equations \eqref{eq:fterm1}, \eqref{eq:fterm2} as the set of polynomials \footnote{Our D--brane system have gauge symmetries which have been  complexified, therefore we have a \textit{projective space} and the homogeneous F--term equations defines a \textit{projective variety}.}.\\
To make progress with \textit{Hilbert series} to be discussed in the last subsection, we recall the definition of an algebraic object, the \textit{ideal} (similar to the idea of a sub--space) such that $\mathcal{I} \subseteq \mathbb{C}\left[x_1, \ldots, x_n\right]$ satisfies the following properties:
\begin{subequations} 
\begin{flalign}
\text{(i)} \, & 0 \in \mathcal{I}. & \\
\text{(ii)} \, & \,\text{If}\, f, g \in \mathcal{I}, \,\text{then}\, f+g \in \mathcal{I}. &\\
\text{(iii)} \, & \,\text{If}\, f \in \mathcal{I} \,\text{and}\, h \in \mathbb{C}\left[x_1, \ldots, x_n\right], \,\text{then}\, h f \in \mathcal{I}. &\\
\text{(iv)} \, & \,\text{Set}\, \left\langle f_1, \ldots, f_s\right\rangle=\left\{\sum_{i=1}^s h_i f_i \mid h_1, \ldots, h_s \in \mathbb{C}\left[x_1, \ldots, x_n\right]\right\} \,\text{is an ideal \footnotemark}\,. &
\end{flalign}
\end{subequations}
\footnotetext{Here, the set is spanned by the generators $f_1, \ldots,f_s$, but we can change basis without affecting the variety.}
By applying the Hilbert Basis theorem \cite{cox1} which in turn relies upon the existence of the \textit{Gr$\ddot{o}$bner  basis} \cite{cox1,buchberger1998grobner} for the ideal $\mathcal{I}$, it can be shown that affine varieties are determined by ideals i.e. $\mathcal{V}(\mathcal{I}) = \mathcal{V}(f_1,\ldots, f_s)$ \footnote{In reverse we have $\left\langle f_1, \ldots, f_s\right\rangle \subseteq \mathcal{I}(\mathcal{V} (f_1,\ldots,f_s))$. But, for algebraically closed fields like $\mathbb{C}$, we can complete the \textit{ideal--variety correspondence} by referring to the much celebrated \textit{Hilbert's Nullstellensatz} \cite{Hilbert1893} which states that $\mathcal{I}(\mathcal{V}(\mathcal{I}))=\sqrt{\mathcal{I}}$ and $\mathcal{V}(\sqrt{\mathcal{I}})=\mathcal{V}(\mathcal{I})$ where $\sqrt{\mathcal{I}}$ is the \textit{radical} ideal.}. \\
We will be mostly interested in the quotients of the polynomial rings, written as a commutative ring $\mathbb{C} [x_1,\ldots,x_n] / \mathcal{I} $, which is the set of equivalence classes for congruence modulo $\mathcal{I}$, i.e.
\begin{equation}
\mathbb{C}\left[x_1, \ldots, x_n\right] / \mathcal{I}=\left\{[g] \mid g \in \mathbb{C} \left[x_1, \ldots, x_n\right]\right\}
\end{equation}
where the equivalence class $[g]$ is 
\begin{equation}
[g]=\left\{f \in \mathbb{C}\left[x_1, \ldots, x_n\right] \mid f - g \in  \mathcal{I}\right\} \,.
\end{equation}
Polynomials with support on the variety $V$ forms a ring $\mathbb{C}[V]$ namely the \textit{coordinate ring} which is isomorphic to $R_V=\mathbb{C}[x_1,\ldots,x_n]/\mathcal{I}(V)$, therefore the ``algebra--geometry" dictionary works for any variety $V$ and $\mathbb{C}[V]$ \cite{cox1}. \\
We will expand on this minimal introduction to computational algebraic geometry as we move along to discuss various methods to efficiently count or in some cases explicitly find solutions to a set of polynomial equations. We shall now discuss four methods in their order of their significance, namely the Newton polytope, Homotopy continuation, Monodromy  and Hilbert series. To put things in perspective, for the first three methods, it is best to think of the fields in our D--brane system as coordinates and the F--term equations as generating a variety. A priori due to gauge symmetries we expect infinite number of solutions as we move along the gauge orbits but upon gauge fixing we expect and get finite number of solutions i.e. zero dimensional varieties with degree equal to the number of solutions \cite{Gray:2008yu}. If this has not been the case, the first three methods would have been ruled out \footnote{As mentioned in the next section \ref{npmethod}, for simpler cases like the case of abelian charges, it is possible to enumerate the independent gauge invariant operators and write the F--term equations in a gauge invariant way. It is increasingly difficult to do so for the non--abelian cases due to the presence of \textit{syzygies}.}. On the contrary, Hilbert series is a more robust quantity which characterizes various important aspects of an algebraic variety. Here, it is best to think of the F--term equations as an ideal modding out the polynomial ring generated by the fields of the D--brane system. We will discuss both gauge fixed and gauge invariant ways of computing the Hilbert series and extract the information about the dimension and degree of the corresponding varieties. 

\subsection{Newton Polytope Method} \label{npmethod}
The study of geometric and combinatorial properties of convex polytopes is an old subject, yet applications of these techniques to questions of mathematical programming is only few decades old \cite{Edelsbrunner1987AlgorithmsIC, Sturmfels1998PolynomialEA}. Here, we are mainly interested in \textit{Newton polytopes} which continue to play a pivotal role in analyzing polynomial systems of equations. We will use two central results for a system of polynomial equations:
\begin{itemize}
    \item \textbf{Bezout's Theorem}: It states that the number of isolated roots of the system is bounded by the product of their degrees \cite{kaveh2008algebraic,Kushnirenko1976NewtonPA}.
    \item \textbf{Bernstein's Theorem}: It states that the number of solutions of a generic system almost equals the \textit{mixed volume} of Newton polytopes corresponding to the polynomial system \cite{Bernshtein1975TheNO, Sturmfels1998PolynomialEA} \footnote{The theorem is for \textit{sparse--square systems} i.e. for equal number of variables and polynomials. Our systems are sparse--square systems. Also, the solutions are assumed to be in the dense tori  $(\mathbb{C}^*)^s$. This is true for abelian charges with $c^{(k\ell)}\neq 0$. For non--abelian case some components of the fields might take zero values at the solutions. To get around it one may use \textit{Cox homotopy} \cite{Duff2020PolyhedralHI} by thinking of the sparse system as a toric variety.}.
\end{itemize} 
Even for the Homotopy continuation method to be discussed in the next section \ref{HomotopyC}, the construction of the start system requires computing mixed volumes of newton polytopes (\textit{polyhedral homotopy}) \cite{Huber1995APM}. We can potentially solve our F--term polynomial equations using this method which we have done for the abelian charges.\\  
A polytope is a subset of $\mathbb{R}^{s}$ that is the convex hull of a finite set of points, $\mathcal{A}=\left\{m_1, \ldots, m_\ell\right\} \subset \mathbb{R}^s$ and expressed as,
\begin{equation}
\operatorname{Conv}(\mathcal{A})=\left\{\lambda_1 m_1+\cdots+\lambda_\ell m_\ell: \lambda_i \geq 0, \, \sum_{i=1}^\ell \lambda_i=1\right\} 
\end{equation}
where, $\operatorname{Conv}(\mathcal{A})$ is the convex hull of $\mathcal{A} \subset \mathbb{R}^s$. A $s$--dimensional polytope has many faces, which are again polytopes of various dimensions between 0 and $(s-1)$. Vertices, edges and facets are the zero, one and $(s-1)$--dimensional faces respectively. 
Consider the following polynomial,
\begin{equation}
    f(x,y)=a_1 x^{u_1}y^{v_1}+a_2 x^{u_2}y^{v_2}+\cdots+a_m x^{u_m}y^{v_m} \, .
\end{equation}
Each monomial $x^{u_i} y^{v_i}$ appearing in $f(x, y)$ corresponds to a lattice point $\left(u_i, v_i\right)$ in the plane $\mathbb{R}^2$. The convex hull of all these points (known as `exponent vectors') is the Newton polytope of $f(x, y)$ \footnote{We always look for the minimal convex hull, some vertices can be on facets or inside the hull. Therefore, the hull has at most $m$ vertices.}, 
\begin{equation}
\operatorname{Newton}(f)=\operatorname{Conv}\left\{\left(u_1, v_1\right),\left(u_2, v_2\right), \ldots,\left(u_m, v_m\right)\right\} \, .
\end{equation}
Another important notion is that of a \textit{Minkowski sum}. If $f_1$ and $f_2$ are any two polytopes in $\mathbb{R}^s$, then their Minkowski sum is the polytope, 
\begin{equation}
f_1+f_2 = \{ p_1 + p_2 \,| \,p_1 \in f_1,\, p_2 \in f_2 \}.
\end{equation}
where each edge of $f_1 +f_2$ is parallel to an edge of either $f_1$ or $f_2$. Geometrically taking the Minkowski sum is mirrored as the algebraic operation of multiplying the polynomials 
\begin{equation}
    \operatorname{Newton}(f_1 \, f_2) =  \operatorname{Newton}(f_1)  + \operatorname{Newton}(f_2) \, .
\end{equation}
The number of solutions in a square system of $s$ polynomials in $s$ unknowns equals the mixed volume $\mathcal{M}$ of the $s$ Newton polytopes. If $f_1, f_2, \ldots, f_s$ are polytopes in $\mathbb{R}^s$ then their mixed volume is  defined by the inclusion--exclusion formula 
\begin{equation}
\mathcal{M}\left(f_1, f_2, \ldots, f_s\right)=\sum_{T \subseteq\{1,2, \ldots,, s\}}(-1)^{s-\#(T)} \cdot \operatorname{Vol} \left(\sum_{s \in T} f_T\right) .
\end{equation}
Equivalently, $\mathcal{M}\left(f_1, f_2, \ldots, f_s\right)$ is the coefficient of the monomial $\lambda_1 \lambda_2 \cdots \lambda_s$ in the expansion of 
\begin{equation}
V\left(\lambda_1, \ldots, \lambda_s\right)=\operatorname{Vol}\left(\lambda_1\, f_1+\lambda_2\, f_2+\cdots+\lambda_s\, f_s\right) \,,
\end{equation}
which is a homogeneous polynomial of degree $s$. Here `$\operatorname{Vol}$' is the usual Euclidean volume in $\mathbb{R}^s$ and $\lambda_1, \ldots, \lambda_s \geq 0$. From Bernstein’s Theorem, the number of solutions to the polynomial equations \eqref{poly} is $\mathcal{M}\left(f_1, f_2, \ldots, f_s\right)$ \footnote{It also works for Laurent polynomials with negative or mixed exponents.}.

\subsubsection{Abelian case}
\begin{figure}[t!]
    \centering
    \begin{subfigure}[t]{0.5\textwidth}
        \centering
        \includegraphics[height=1.2in]{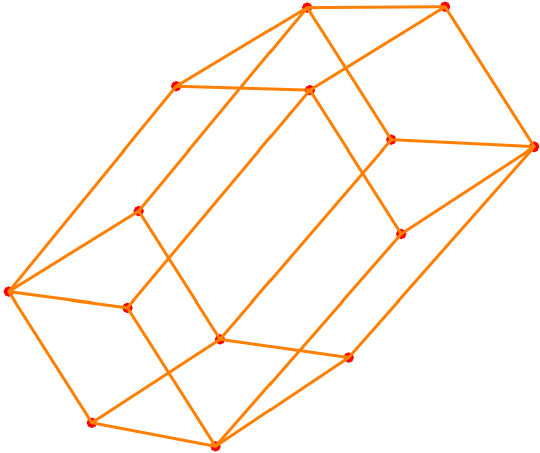}
        \caption{Newton Polytope for $f_1+f_2$ \\
                   (14 vertices)}
    \end{subfigure}%
    ~ 
    \begin{subfigure}[t]{0.5\textwidth}
        \centering
        \includegraphics[height=1.2in]{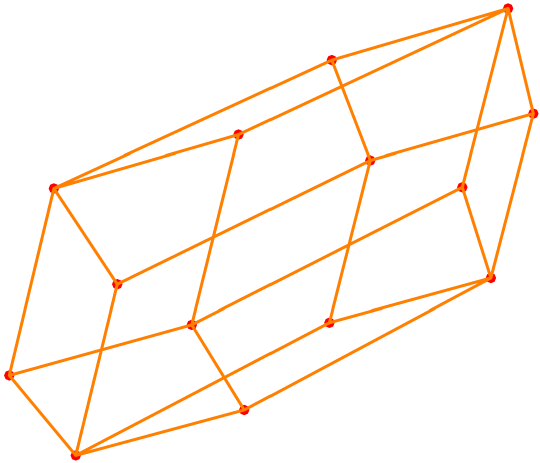}
        \caption{Newton Polytope for $f_1+f_3$ \\
                   (14 vertices)}
    \end{subfigure}
    ~
    \begin{subfigure}[t]{0.5\textwidth}
        \centering
        \includegraphics[height=1.2in]{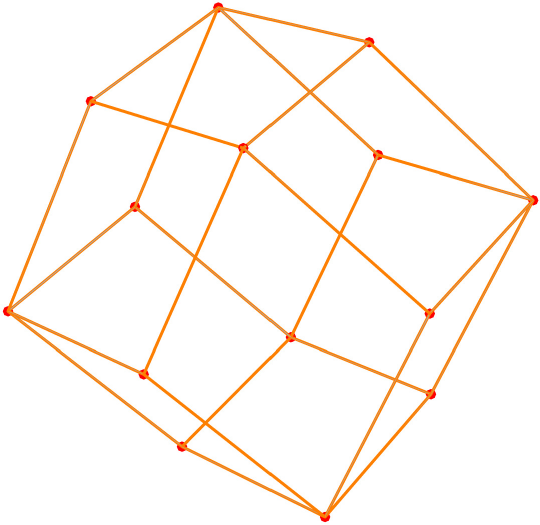}
        \caption{Newton Polytope for $f_2+f_3$ \\
                   (14 vertices)}
    \end{subfigure}%
    ~ 
    \begin{subfigure}[t]{0.5\textwidth}
        \centering
        \includegraphics[height=1.2in]{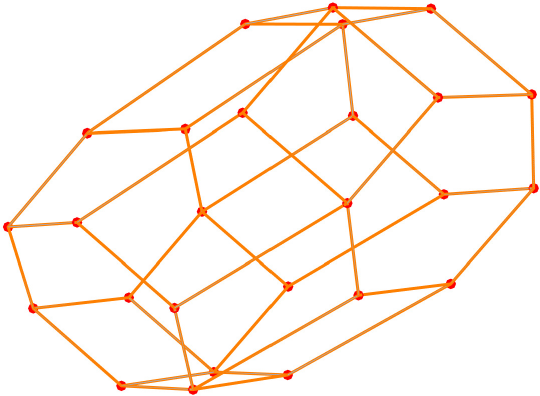}
        \caption{Newton Polytope for $f_1+f_2+f_3$\\
                   (26 vertices)}
    \end{subfigure}

    \caption{Newton Polytopes for charge $(1,1,1,1)\,$.}
\label{fig:polytope}
   
\end{figure}

We now apply the above method to solve the system of F--term equations coming from \eqref{eq:fterm2} after eliminating the $\Phi$ fields. The system of equations are 
\begin{equation}\label{abelianf}
\begin{aligned}
& Z^{(23)} Z^{(31)} Z^{(12)}+Z^{(23)} Z^{(34)} Z^{(42)}=Z^{(32)} Z^{(21)} Z^{(13)}+Z^{(32)} Z^{(24)} Z^{(43)} \\
& Z^{(24)} Z^{(41)} Z^{(12)}+Z^{(24)} Z^{(43)} Z^{(32)}=Z^{(42)} Z^{(21)} Z^{(14)}+Z^{(42)} Z^{(23)} Z^{(34)} \\
& Z^{(34)} Z^{(41)} Z^{(13)}+Z^{(34)} Z^{(42)} Z^{(23)}=Z^{(43)} Z^{(31)} Z^{(14)}+Z^{(43)} Z^{(32)} Z^{(24)}
\end{aligned}
\end{equation}
and 
\begin{equation}
Z^{(k\ell)}Z^{(\ell K)}=-c^{(k \ell)} \quad \text { for } \quad 1 \leq k<\ell \leq 4 \, .
\end{equation}
We can define gauge invariant variables $u$'s, invariant under the $U(1) \times U(1) \times U(1)$ gauge transformations as 
\begin{equation}
\begin{aligned}
& u_1 \equiv Z^{(12)} Z^{(21)}, \quad u_2 \equiv Z^{(23)} Z^{(32)}, \quad u_3 \equiv Z^{(31)} Z^{(13)} \\
& u_4 \equiv Z^{(14)} Z^{(41)}, \quad u_5 \equiv Z^{(24)} Z^{(42)}, \quad u_6 \equiv Z^{(34)} Z^{(43)} \text {, } \\
& u_7 \equiv Z^{(12)} Z^{(24)} Z^{(41)}, \quad u_8 \equiv Z^{(13)} Z^{(34)} Z^{(41)}, \quad u_9 \equiv Z^{(23)} Z^{(34)} Z^{(42)}\, . 
\end{aligned} 
\end{equation}
The variables $(u_1,\ldots, u_6)$ simply take the values $- c^{(k\ell)}$ and the remaining equations \eqref{abelianf} can be written as
\begin{equation}\label{eq:u}
\begin{aligned}
a_{1} u_7^{2} u_9^{2}-a_{2} u_7 u_8-a_{3} u_8^{2}-u_7 u_8 u_9^{2} &= 0 \\
u_7^{2} u_9- u_7 u_9^{2}-a_{4} u_9+a_{5} u_7 &=0 \\
u_8^{2} u_9-a_{6} u_8-a_{7} u_9+u_8 u_9^{2}  &=0 \, ,
\end{aligned}
\end{equation}  
where $a_i$ are generic functions of the tunable constants $c^{(k\ell)}$. To apply the Newton polytope method, it is important that  $a_i$'s are generic and  $a_i\neq 0$. \\
The exponent vectors of the above three polynomials $(f_1, \, f_2, \, f_3)$ are $\{(2,0,2),\, (1,1,0),\,\\ (0,2,0), (1,1,2)\}$, $\{(2,0,1),\, (1,0,2),\, (0,0,1),\, (1,0,0)\}$ and $\{(0,2,1),\, (0,1,0),\, (0,0,1),\,\\ (0,1,2)\}$ respectively, see figure \ref{fig:polytope}. The number of solution to these three polynomial equations will be the mixed volume of $f_1$, $f_2$ and $f_3$, i.e.
\begin{align}
\label{mv123}
    \mathcal{M}\left(f_1, \,f_2,\, f_3\right)= & \operatorname{Vol}(f_1 +f_2+ f_3)-\left \{\operatorname{Vol}(f_1+f_2)+\operatorname{Vol}(f_1+f_3)+\operatorname{Vol}(f_2+f_3)\right \} \notag \\
    & +\left \{\operatorname{Vol}(f_1)+\operatorname{Vol}(f_2)+\operatorname{Vol}(f_3) \right \} \, .
\end{align}
It turns out that $f_1$, $f_2$ and $f_3$ are 2--dimensional polytopes in $\mathbb{R}^3$, hence they have zero volume. Remaining volumes are non zero and has been computed by writing a code in Mathematica \cite{Mathematica} and by using software package SAGE \cite{sagemath}. The values of the volumes are as below, 
\begin{equation}
  \operatorname{Vol}(f_1+f_2)= \operatorname{Vol}(f_1+f_3)=\operatorname{Vol}(f_2+f_3)= 8, \, \operatorname{Vol}(f_1+f_2+f_3)=36 \, 
\end{equation}
and $\mathcal{M}\left(f_1, \,f_2,\, f_3\right)=12$, which matches with earlier results of microstate counting \cite{Chowdhury:2014yca,Chowdhury:2015gbk}. We can start with the gauge fixed version of the abelian charges and repeat the calculations with more sophisticated algorithms using \textit{mixed subdivision} of newton polytopes \cite{Li2008NumericalSO, Huber1995APM}, the result remains the same. 

\subsubsection{Non--abelian cases}
In principle, by applying the above method we should be able to obtain the correct number of solutions to the gauge fixed F--term equations, but there are caveats.
\begin{itemize}
    \item As we increase the charges $(N)$, the number of variables grows  as $\mathcal{O}(N^2)$. Newton polytope method is a $\mathbb{P}$--hard problem \cite{10.1137/S0097539794278384}, so computations will take forever \footnote{The number of $\operatorname{Vol}$ computations scales roughly as $2^{N^2} =e^{N^2 \ln 2}$. So, complexity scales as $2^{N^2} \times$ $\mathbb{P}$--hardness of Convex hull volume computation.}.
    \item The solutions are supposed to be on a dense tori $(\mathbb{C}^*)^s$ but for non--abelian cases some field components could take values in $\mathbb{C}$ i.e. be zero.
    \item Most importantly, the method is valid for generic coefficients of the polynomials and for non--abelian cases the coefficients of our F--term equations are not generic.  
\end{itemize}
Let us expand on the last point. In \cite{Chowdhury:2015gbk}, it was shown that if we choose the superpotential term $\mathcal{W}_4$ as 
\begin{align}
\mathcal{W}_4  = & -\sqrt{2}\, C \left[\operatorname{Tr}\left(\Phi_1^{(1)} \Phi_2^{(1)} \Phi_3^{(1)}-\Phi_1^{(1)} \Phi_3^{(1)} \Phi_2^{(1)}\right)-\operatorname{Tr}\left(\Phi_1^{(2)} \Phi_2^{(2)} \Phi_3^{(2)}-\Phi_1^{(2)} \Phi_3^{(2)} \Phi_2^{(2)}\right)\right.\notag \\
& \left.+\operatorname{Tr}\left(\Phi_1^{(3)} \Phi_2^{(3)} \Phi_3^{(3)}-\Phi_1^{(3)} \Phi_3^{(3)} \Phi_2^{(3)}\right)+\operatorname{Tr}\left(\Phi_1^{(4)} \Phi_2^{(4)} \Phi_3^{(4)}-\Phi_1^{(4)} \Phi_3^{(4)} \Phi_2^{(4)}\right)\right]
\end{align}
with $C=1$, we get the correct count 56 for charges $(1,1,1,2)$ and 208 for charges $(1,1,1,3)$. If we choose $C\neq1$, we get wrong counts or if we choose $\mathcal{W}_4$ to be 
\begin{align}
\mathcal{W}_4= & -\sqrt{2}\left[\operatorname{Tr}\left(\Phi_1^{(1)} \Phi_2^{(1)} \Phi_3^{(1)}-\Phi_1^{(1)} \Phi_3^{(1)} \Phi_2^{(1)}\right)+\operatorname{Tr}\left(\Phi_1^{(2)} \Phi_2^{(2)} \Phi_3^{(2)}-\Phi_1^{(2)} \Phi_3^{(2)} \Phi_2^{(2)}\right)\right. \notag \\
& \left.+\operatorname{Tr}\left(\Phi_1^{(3)} \Phi_2^{(3)} \Phi_3^{(3)}-\Phi_1^{(3)} \Phi_3^{(3)} \Phi_2^{(3)}\right)+\operatorname{Tr}\left(\Phi_1^{(4)} \Phi_2^{(4)} \Phi_3^{(4)}-\Phi_1^{(4)} \Phi_3^{(4)} \Phi_2^{(4)}\right)\right]\, ,
\end{align}
we get 60 for charges $(1,1,1,2)$ and 232 for charges $(1,1,1,3)$, again wrong counts. Therefore, the polynomials of our system are not generic and the answer depends on the specific coefficients. Though, we will not pursue it any further, we have explained the method in some detail as it forms the basis for Homotopy continuation method discussed in the next section.

\subsection{Homotopy Continuation Method}\label{HomotopyC}
A fundamental tool in computational algebraic geometry is homotopy continuation, which computes the isolated solutions to a polynomial system $F=\{f_1,\ldots,f_s\}$ by numerically tracking paths that interpolate between the solutions of $F$ and an already known solutions of a similar system $G=\{g_1,\ldots,g_s\}$ \cite{Hauenstein2017WhatIN, Hauenstein2022ApplicationsON,  Sommese2005TheNS}. For the purposes of this paper we assume both $F$ and $G$ are square sparse systems that define zero dimensional varieties \footnote{Both over--determined and under--determined systems, as well as projective varieties can be solved.}. We have used a modified version of HomotopyContinuation.jl \cite{HomotopyContinuation}, a Julia package designed for this purpose to solve the F--term equations for charges $(1,1,1,1)$, $(1,1,1,2)$, $(1,1,1,3)$ and the previously unreported $(1,1,1,4)$. In all these cases, the supersymmetric vacua count matches with the numbers coming from the U--dual system, appendix \ref{14helicitytrace}. \\
 We begin by defining $H(x ; \tau) \subset$ $\mathbb{C}\left[x_1, \ldots, x_s\right][\tau]$ as a one--parameter family of equations with parameter $\tau \in \mathbb{C}_\tau$ such that both $H(x ; 0)=G$  and $H(x ; 1)=F$ belong to the family. Then $\mathcal{V}(H) \subset \mathbb{C}^s \times \mathbb{C}_\tau$ is a complex family of varieties realized as a fibre bundle by the projection $\pi: \mathcal{V}(H) \rightarrow \mathbb{C}_\tau$ i.e. $\mathcal{V}(G)$ in the fiber over $\tau=0$ and $\mathcal{V}(F)$ over $\tau=1\,$. 
 Let $C$ denote the union of all the one--dimensional components of $\mathcal{V}(H)$ such that its restriction, $\left.C\right|_\gamma\,$, to an arc $\gamma \subset \mathbb{C}_\tau$ between 0 and 1 is smooth and allows for tracking of every point in $\mathcal{V}(G)$ at $\tau=0$ to a point in $\mathcal{V}(F)$ at $\tau=1$ \footnote{This is possible if a point $(x ; \tau) \in \mathcal{V}(H)$ is non--degenerate i.e. the Jacobian of $H$ with respect to the $x$--variables, $D_x H$, is invertible at $(x ; \tau)\,$.}. An example of a simple homotopy is the \textit{straight--line homotopy}, which is defined by $H(x ; \tau)=(1-\tau)\,G+\tau\,F $. 

 \subsubsection*{Path tracking}
 
 \begin{wrapfigure}{r}{0.50\textwidth}
 \vspace{-0.5cm}
 \begin{framed}
 \centering
\includegraphics[scale=0.7]{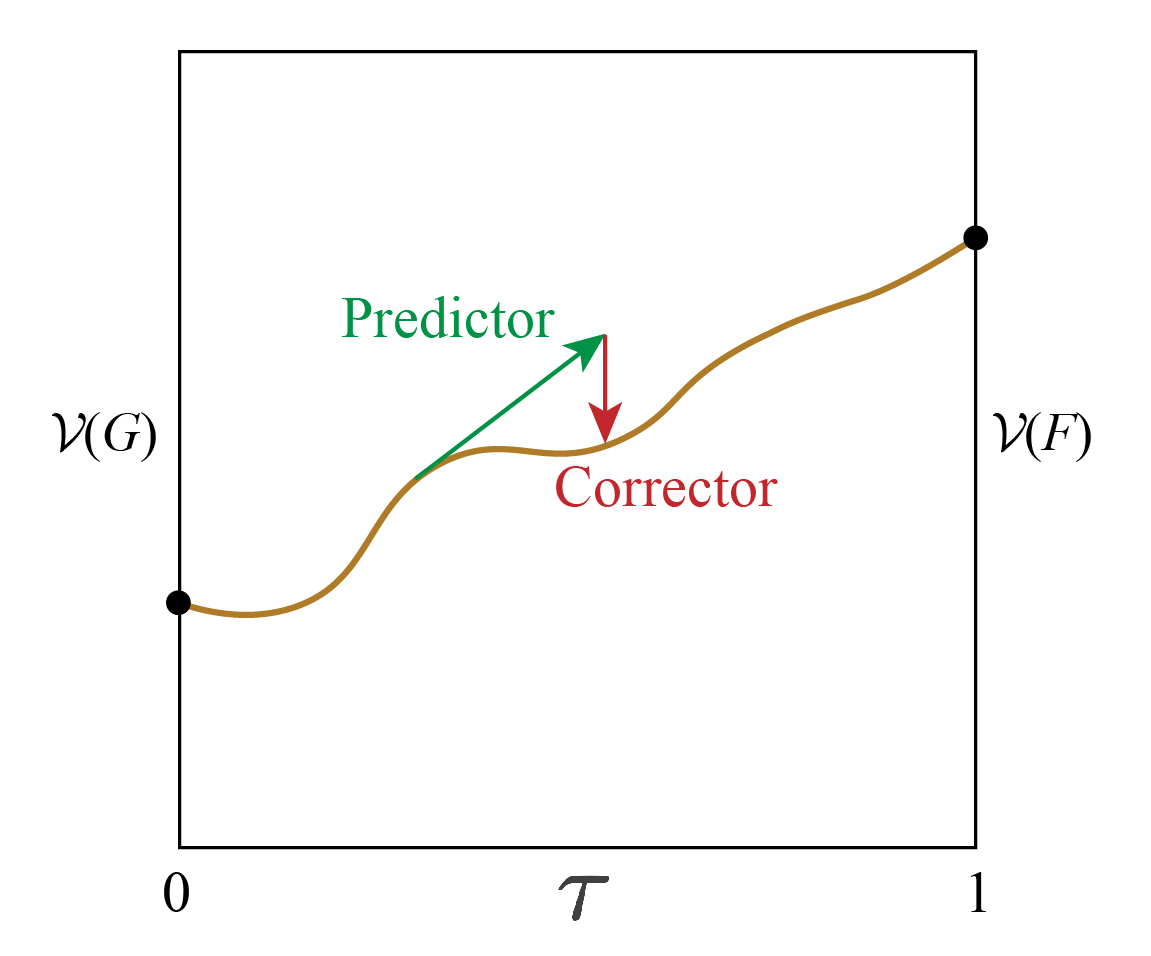} 
\caption{Schematic of path tracking, showing prediction and correction steps.} 
\label{fig:path}
\end{framed}
\vspace{-1.0cm}
\end{wrapfigure}
 Path tracking amounts to numerically solving an initial value problem. For every solution, by the Implicit Function Theorem, there is a parametrization $x(\tau)$ for each arc such that $H(x(\tau) ; \tau)=0$ and $x(0)=x_0$ is known.  Differentiating $H(x(\tau) ; \tau)=0$ with respect to $\tau$ gives a Davidenko differential equation \cite{davidenko1953new} as a initial value problem,
 \begin{equation}
 \left\{\begin{array}{l}
\displaystyle \frac{d x}{d \tau}=-\left(D_x H\right)^{-1} \frac{\partial H}{\partial \tau}\vspace{0.25cm}\\
x(0)=x_0 .
\end{array}\right.
\end{equation}
which can be solved by applying any numerical differential equation solver. To maintain numerical stability, it is best to use the \textit{predictor--corrector} algorithm \footnote{Starting from a point close to the solution curve, \textit{Euler Forward method} (predictor) generates a solution one step further, then  \textit{Newton's method} of root approximation (corrector) is used to refine the solution at that step and the iterations continues till $x(\tau)$ is found at $\tau=1$. Path tracking is reasonably straightforward to parallelize and can be further improved by the use of GPU \cite{DBLP:journals/corr/abs-2112-03444}.} \cite{allgower2003introduction}, see Figure \ref{fig:path}. Path tracking rely on inverting $D_x H$, and will fail close to $\tau=1$ if $\mathcal{V}(F)$ contains singular solutions. In such cases tracking is stopped very close to $\tau=1$ and an \textit{endgame} algorithm like \textit{Cauchy endgame} \cite{Bates2010APE} takes  over.

\subsubsection*{Start system}

The choice of start system $G$ and homotopy $H$ is important to maintain the efficacy and efficiency of homotopy algorithms. The start system $G$ should have at least the same number of solutions as $F$ and if it has more solutions then the extra  paths diverge to infinity \footnote{This happens for example in \textit{Bézout} or \textit{total--degree homotopy}, where $G=\{x_{1}^{d_1}-1= 0,\, x_{2}^{d_2}-1=0,\cdots,x_{s}^{d_s}-1=0\}$ and $d_i=\operatorname{deg}(f_i)$. Usually, he Bézout bound $d_1d_2\ldots d_s \gg \operatorname{deg}(\mathcal{V}(F))$.}. The setup is optimal when $F$ and $G$ in homotopy $H(x ; \tau)$ have the same number of solutions when counted with multiplicity, and if the paths connecting $\mathcal{V}(G)$ and $\mathcal{V}(F)$ are smooth everywhere except possibly at $\tau=1$. The algorithms we have used relied on polyhedral homotopy \cite{Huber1995APM} which used \textit{toric degenerations} \cite{coxhomotopy} such that the start system, $\mathcal{V}(G)$, is a toric variety and used tropical geometry to enumerate the mixed cells of a mixed subdivisions \cite{Jensen2016TropicalHC}. 

\subsubsection*{Our strategy}
The F--term equations \eqref{eq:fterm1}, \eqref{eq:fterm2} for both the abelian and non--abelian cases are not a sparse square system; not all of them are independent. We can convert it to a square system by appropriately fixing the gauge and using the phase symmetries \eqref{eq:phase}, along with the removal of the corresponding F--term equations from the polynomial equations list. In all cases, some of the remaining F--term equations now become linear, we can use this opportunity to eliminate them and reduce the system a bit further. With these reduced set of variables and F--term equations forming a square system, we discuss below the cases of our interest.

\subsubsection{Abelian case}
\begin{wraptable}{r}{8cm}
\vspace{-0.65cm}
\begin{framed}
\centering
   $
   \begin{array}{|c|c|}
    \hline Z^{(12)}&1 \\
    \hline Z^{(34)}&1 \\
    \hline Z^{(13)}&1 \\
    \hline Z^{(21)}&-1.11\\
    \hline Z^{(31)}& -1.13\\
    \hline Z^{(14)}&0.589628+ 0.0708712 i \\
    \hline Z^{(41)}&-1.90589+ 0.229081 i\\
    \hline Z^{(23)}&-0.424973- 0.792931 i\\
    \hline Z^{(32)}&0.588095 -1.09729 i\\
    \hline Z^{(24)}& 0.41384 - 0.601617 i\\
    \hline Z^{(42)}&-0.892554- 1.29754 i\\
    \hline Z^{(43)}& -1.16\\
    \hline \Phi^{(2)}_{3}& 0.153779 - 1.92561i \\
    \hline \Phi^{(3)}_{1}&0.151189 - 1.92426 i\\
    \hline \Phi^{(3)}_{2}&1.10894 + 0.875142i  \\
    \hline \Phi^{(4)}_{1}&-1.10085- 0.888016 i\\
    \hline \Phi^{(4)}_{2}&0.145618 - 1.89443 i\\
    \hline \Phi^{(4)}_{3}& -1.08305- 0.887995 i \\
    \hline
   \end{array}
   $
    \caption{One of the 12 SUSY vacuum for charges $(1,1,1,1)$. First three entries denote the gauge fixing.}
    \label{hessian}
    \end{framed}
\end{wraptable}
After fixing the gauge, the F--term equations become a square system of 15 polynomials in 15 variables. With parallel path tracking, 16 paths were tracked very quickly (seconds) to give 12 non--singular solutions which of course matches with the expectation. Here, we would like to discuss further two points related to the Hessian and the D--terms which are also valid for the non--abelian cases. \\
The real Hessian $h$ around all explicit solutions for both abelian and non--abelian cases can be computed. It is also possible to diagonalize the Hessian and look at it eigenvalues and eigenvectors \footnote{For non--abelian cases, even with large number of variables, standard routines in python will do the job.}. For concreteness, we look at eigenvalues of the real Hessian for a particular solution of the abelian charges, Table \ref{hessian} with $C=1$, $c^{(12)}=1.11$, $c^{(23)}=1.12$, $c^{(13)}=1.13$, $c^{(14)}=1.14$, $c^{(24)}=1.15$ and $c^{(34)}=1.16\,$. The list of eigenvalues are $\{113.174,\,-113.174,\,113.171,\,-113.171,\,112.223,
\,-112.223,\,-112.197, \,112.197, -112.052,\,\\ 112.052, \,112.03,\,-112.03,\,1.90029,\,-1.90029,\, 1.89812,\,-1.89812,\,-1.15207,\, 1.15207,\,\\ -1.15027,\,1.15027, \,
1.02881,\,-1.02881,\,-1.01489,\,1.01489,\,-0.852902,0.852902,\,-0.841017,\,\\0.841017,\, -0.00169278,\,0.00169278,\,0,\,0,\,0,\,0,\,0,\,0 \}$. As discussed in section \ref{susyvacua}, there are equal numbers of positive and negative eigenvalues. The six zero eigenvalues are the massless Goldstone modes coming from gauge fixing the global complexified gauge $U(1) \times U(1) \times U(1)$. The same can be verified for the other 11 solutions and also for all solutions of the non--abelian cases. \\
 Starting with a particular gauge fixed solution of the F--term equations, we can move along the complex gauge orbits to generate an infinite set of equivalent F--term solutions. Picking one such solution from the orbit will not automatically solve the D--term equations which are invariant under the physical gauge transformations. Therefore, for both abelian and non--abelian cases, our F--terms solutions don't satisfy the D--term equations and the potential $V_D$ \eqref{dtermpotential} evaluates to a non zero number. To remedy this, for each solution, we retain the `physical gauge fixing' and set up a minimization problem of $V_D=0$ with the additional `complex' part of the gauge group as parameters. Upto physical gauge transformations, in all cases we find unique solutions which satisfy both the F--term and D--term equations simultaneously \footnote{As an illustration, take the abelian case where the superpotential is invariant under the complexified $U(1)$ gauge transformations $Z^{(k \ell)} \rightarrow a_k\left(a_{\ell}\right)^{-1} Z^{(k \ell)}$ for $1 \leq k,\, \ell \leq 4, \, k \neq \ell$ and all $\Phi$ fields are gauge invariant. We can set $a_4=1$ to mod out the overall diagonal $U(1)$. Moving along the gauge orbits, suppose we find on set of $a_1$, $a_2$ and $a_3$ such that the D--terms are satisfied. The physical gauge transformation $a_k \rightarrow e^{i \phi_k} a_k$ also satisfies the D--term equations. Therefore, using the transformations generated by the $\phi_k$'s we can fix the `gauge' for $a_1, a_2$ and $a_3$ by setting $a_k$'s to be real and positive and then minimize $V_D$ over these parameters.}.   

\subsubsection{Non--abelian cases}

We report here the results of applying the homotopy continuation method to F--term equations of non--abelian charges reduced to sparse square systems.
\begin{itemize}
    \item \textbf{Charge $(1,1,1,2)$}: We have 24 variable and 24 polynomials. As expected, tracking of about 297 paths resulted in 56 non--singular solutions.
    \item \textbf{Charge $(1,1,1,3)$}: We have 32 variable and 32 polynomials. As expected, tracking of about 3736 paths resulting in 208 non--singular solutions \footnote{It should be noted that it took only about a minute in our PC while Mathematica latest version took more than 5 hours. For charges $(1,1,1,4)$ Mathematica failed but homotopy took a little more than two hours.}. 
    \item \textbf{Charge $(1,1,1,4)$}: We have 47 variables and 47 polynomials. As expected, tracking of about 2.5 lakh paths resulting in 684 non--singular solutions \footnote{Single runs resulted in $680\pm 2$ solutions. Union of the solutions from two different runs with different \textit{seeds}, always resulted in 684 unique solutions.}.
\end{itemize}
It is clear from these numbers that complexity is increasing at a fast pace. Again it is the mixed cell computations that are $\mathbb{P}$--hard but a few more higher charges can be tried if one is willing to wait for longer times. In the next section we will describe an arguably better way to construct the start systems.

\subsection{Monodromy  Method} \label{monodromy}
In the previous subsection on Homotopy continuation, the main bottleneck has been the construction of a generic start system $G$ such that $\operatorname{deg}(\mathcal{V}(G))\geq \operatorname{deg}(\mathcal{V}(F))$. As an alternative and often faster route, we could exploit the action of a \textit{monodromy group} to explore the variety $\mathcal{V}(F)$ \cite{duffmonodromy, Bliss2018MonodromySS}. We begin by deforming the sparse square system $F$ to a $k$--parameter family of polynomials, $F(x ; p): \mathbb{C}^s \times \mathbb{C}^k \rightarrow \mathbb{C}^s$, such that 
\begin{equation}
F(x, p) =\left[\begin{array}{c}
f_1\left(x_1, \ldots, x_s ; p_1, \ldots, p_k\right) \\
\vdots \\
f_s\left(x_1, \ldots, x_s ; p_1, \ldots, p_k\right)
\end{array}\right]=0 \,,
\end{equation}
For generic choices of parameters $p\in \mathbb{C}^k$ it is expected that isolated solutions of $F(x;p)$ will remain constant. Hence, by using \textit{parameter homotopy} $\displaystyle H(x, \tau)=F(x, \gamma(\tau))=0$ \cite{Sommese2005TheNS}, specific solutions for parameters $p$ can be tracked to other parameter values $p^{\prime}$ by following continuous paths $\gamma(\tau):[0,1] \rightarrow \mathbb{C}^k$ such that $\gamma(0)=p$ and $\gamma(1)=p^{\prime}$. To see what happens if $\gamma$ is chosen to be closed path, we notice that the projection $\pi: \mathcal{V}(F(x;p)) \rightarrow \mathbb{C}^k$ defined by $(x, p) \mapsto p$, has an open set $U \subset \mathbb{C}^k$ where the restriction of $\pi$ onto $U$ is generically a finite morphism of degree $d=\operatorname{deg}(\mathcal{V}(F))$.  A loop in $U$ based at $p$ has $d$ lifts to $\pi^{-1}(U)$, one for each point in the fiber $\pi^{-1}(p)$. The monodromy group of the map $\pi$ is the image of the usual permutation action of the fundamental group of $U$ in the space of solutions $S_d$. Therefore, if one solution for some $p^* \in \mathbb{C}^k$ is known, then we track the solution $x^* \in \mathbb{C}^s$ to a new solution $x^{\prime} \in \mathbb{C}^s$ of $F(x;p)$ by following a closed path (loop) in parameter space such that $\gamma(0)=p^* =\gamma(1)$. If $S_r \subseteq S_d$ is a set of solutions obtained by repeating this process $r$ times, we obtain a new solution set by taking a new random loop to track $S_r$ to $S_r^{\prime}$ and letting $S_{r+1}=S_r \cup S_r^{\prime}$. Notice that $S_r^{\prime}$ may coincide with $S_r$, but we always have the inclusions $S_r \subseteq S_{r+1} \subseteq S_d$. We continue constructing random monodromy paths until $S_{r+1}=S_d$.   

\subsubsection*{Our Strategy}
We make the following parameter deformation on our F--term equations
\begin{equation}
F(x, p) =\left[\begin{array}{c}
f_1(x_1, \ldots, x_s)-p_1 \\
\vdots \\
f_s(x_1, \ldots, x_s)-p_s
\end{array}\right]=0 \,.
\end{equation}
such that it is trivial to find a solution for $p^*$ for any random values for $x^*$. We take one such pair $(x^*,p^*)$ as the seed solution of $F(x;p)$ and then carry out the monodromy anchored at $p^*$. Once we have the set of solutions for $F(x;p^*)=0$, we run a parameter homotopy on this set tracking solutions from $\gamma(0)=p^*$ to $\gamma(1)=0$, thereby generating the final set of solutions for $F(x;0)$ \footnote{Alternatively, we could have performed the parameter homotopy at the beginning to go from the solution $F(x^*,p^*)$ to $F(x^{\prime},0)$ and then performed the monodromy around $p=0$ to generate all solution of $F(x;0)$.}. The computational advantage of the monodromy method is the parallelizability of both the monodromy computations and the path tracking  of parameter homotopy.

\subsubsection*{Results}

The results for the abelian charge and non--abelian charges up to $(1,1,1,3)$ are as expected. For charges $(1,1,1,4)$ it missed two solutions on single run but union of two different runs give 684 solutions. On an average the monodromy method is about ten times faster than the polyhedral homotopy discussed in the previous subsection. We believe this method requires further attention primarily because of two reasons
\begin{itemize}
    \item We can $k$--parameterize the F--term equations in different ways. Here, we took the most straightforward one.
    \item The solution sets have discrete symmetries, for example a discrete $\mathbb{Z}_2$ coming from complex conjugation of all fields. It should be possible to incorporate them automatically while populating the fibres $\pi^{-1}(p^*)$ or $\pi^{-1}(0)$. 
\end{itemize}
 Its original implementation in terms of minimizing a \textit{cost function} also deserves further attention  \cite{Campo2015CriticalPV}. We are currently looking into these points to further boost efficiency. Recently, the efficacy of \textit{Stochastic Gradient Descent} \cite{EonBottou1998OnlineLA} in minimizing the square potential $V_F$ for $\mathcal{N}=8$ black holes is discussed in \cite{Kumar:2023hlu} but results don't match beyond charges $(1,1,1,2)$. We believe that in applications to machine learning, where the \text{loss function} in many neural network architectures, have exponentially many low lying local minima \cite{Auer1995ExponentiallyML}, more sophisticated techniques like the current method will do better.\\ We will now move from numerical solutions to exact (symbolic) techniques in the next sub--section, where similar to the Newton polytope method discussed at the beginning of this section we will just count the number of solutions to the F--term equations without ever having to know the solutions explicitly.

\subsection{Hilbert Series}
So far we have discussed three methods to deal with solving polynomial equations but they have their limitations, the biggest of all being limited to mostly zero dimensional varieties. We shall now discuss a very important quantity characterizing an algebraic variety, namely the Hilbert series. We have used it in a limited capacity to count the number of gauge invariant operators which tells us about the dimension and degree of the variety -- the vacua manifold. In our cases,  the vacua manifolds are essentially the `Higgs branch' that too zero dimensional. We believe that moving away from the generic background metric and $B$--field moduli space will result in more complicated structures for the vacua manifolds but hopefully the Witten index will remain the same. In such cases the machinery developed by Hanany and friends \cite{Hanany_adjointsqcd,Hanany:baryonicgeneratingfn,Hanany:chiraloperatorquiver,Hanany:bpsoperators,Hanany:plethystic,Gray:2008yu} applying Hilbert series techniques to SQCD's, Quiver gauge theories, Conifolds etc. to probe the structure of the vacua manifolds will be useful. We will venture along these lines in future works.\\
One of the key insights of Hilbert in \cite{Hilbert1890} was that 
 the dimension of a variety associated to an ideal could be characterized by the growth of the number of polynomials not in the ideal as the total degree increases. This later formed the basis for the Hilbert functions, Hilbert polynomials and subsequently the Hilbert series. The Hilbert series of a variety is not a topological quantity as it depends on the embedding $\mathbb{C}^n$ for the variety. A specific embedding then induces a grading on the coordinate ring, $\displaystyle R_V=\bigoplus_{i \geq 0} R_i$, where $R_i$ is the collection of independent degree $i$ polynomials on the variety $V$. The Hilbert series is the generating function for the dimensions of the $\mathbb{C}$--vector space spanned by these graded pieces of $R_V$,
\begin{equation} \label{hs1}
H(t ; V)=\sum_{i=0}^{\infty}\left(\operatorname{dim}_{\mathbb{C}} R_i\right) t^i \, .
\end{equation}
Here, $t$ is a dummy variable and it is called `fugacity' much like the `chemical potential' in statistical mechanics. Hilbert series is a rational function in $t$ which can be written in two different ways \cite{Atiyah1969IntroductionTC}: 
\begin{equation} \label{HS2}
H(t ; V)= \begin{cases}\frac{Q(t)}{(1-t)^n} & \text { Hilbert series of First kind } \\ \frac{P(t)}{(1-t)^{\operatorname{dim}(V)}} & \text { Hilbert series of Second kind }\end{cases}
\end{equation}
where $P(t)$ and $Q(t)$ are polynomials with integer coefficients. The powers of the denominators are such that the leading pole captures the dimension of the embedding space and the variety $V$, respectively.
In particular, $P(1)$ always equals the degree of the variety which for the case of zero dimensional varieties is the number of solutions to the polynomial equations \cite{Gray:2008yu}.\\
For a $n$--dimensional  variety $V$, a Laurent series expansion for the Hilbert series of second kind in \eqref{HS2}  can be developed as a partial fraction expansion:

\begin{equation}
H(t, V)=\frac{V_n}{(1-t)^n}+\cdots \frac{V_3}{(1-t)^3}+\frac{V_2}{(1-t)^2}+\frac{V_1}{1-t}+V_0+\mathcal{O}(1-t)\,,
\end{equation}
where we can see explicitly that the Hilbert series is a rational function and the degree of its most singular pole is the dimension of $V$.\\
It is possible to generalize the grading to a $m$--multiple grading of the coordinate ring $R_V$ \cite{MultigradedHF}, such that $\displaystyle R_V=\bigoplus_{i_1, i_2, \ldots, i_m \geq 0} R_{i_1, i_2, \ldots, i_m}$, and the corresponding Hilbert series is
\begin{equation}
H\left(t_1, t_2, \ldots, t_m ; V \right)=\sum_{i_1, i_2, \ldots, i_m=0}^{\infty}\left(\operatorname{dim}_{\mathbb{C}} R_{i_1, i_2, \ldots, i_m}\right) t_1^{i_1} t_2^{i_2} \ldots t_m^{i_m} \, .
\end{equation}
We will use the \textit{multi--graded} Hilbert series in extracting the count of gauge invariant operators.

\subsubsection*{Our Strategy}
We construct the Hilbert series from the F--term constraints as follows \cite{jejjala:2006jb} : 
\begin{enumerate}
\item We start with the polynomial ring $\mathbb{C}\left[x_1,..x_n\right]$ of the basic variables $Z^{(k \ell)}$ and $\Phi^{(k)}_i$ (which are not gauge invariant). 
\item Impose $F$--term equations as an ideal $\mathcal{I}$ on this ring and compute the quotient ring $\mathbb{C}\left[x_1,..x_n\right]/\mathcal{I}$ \footnote{In practice, in several situations we have replaced the ideal $\mathcal{I}$ by its standard or reduced Gröbner basis ideal. These Gröbner bases has been enumerated using a parallel implementation of the \textit{F4 algorithm} as discussed in \cite{Monagan2015ACP}. }. 
\item Compute the Hilbert series and extract the gauge invariant part of $\mathbb{C}\left[x_1,..x_n\right]/\mathcal{I}$, which would then be identified with the coordinate ring of the vacua manifold as we expect it to be spanned by gauge invariant operators \footnote{We could have started with a more satisfying approach of writing everything in terms of gauge invariant coordinates. The gauge invariant F--term equations are generators of an ideal of the ring of invariants. For the abelian charges we can do it in terms of the $u$ variables discussed in \ref{npmethod}. However for the non--abelian cases, finding generators of the ring of invariants is a difficult task due to presence of hidden relations called \textit{syzygies}. For example, take the case of a three node abelian quiver with bi--fundamental fields $z_{12}, \,z_{21}, \, z_{13}, \,z_{31}, \,z_{23}, \, z_{32}$. Naively, the gauge invariant coordinates would be $x_1=z_{12} z_{21}, \, x_2=z_{13} z_{31},\, x_3=z_{23} z_{32},\, x_4=z_{12} z_{23} z_{31}$ and $ x_5=z_{13} z_{32} z_{21}$, but they are not all independent. There is a relation, the syzygy $x_1 x_2 x_3=x_4 x_5\,$. Hence even without any F--terms the coordinate ring is $\displaystyle \mathbb{C}\left[x_1, x_2, x_3, x_4, x_5\right]/\left(x_1 x_2 x_3-x_4 x_5\right)$. To the best of our knowledge, there is no known algorithm to lay out the set of genuinely independent gauge invariant coordinates.}.
\end{enumerate}
We will now expand on the grading and extraction part of the procedure.

\subsubsection*{Grading and Extraction}

We endow appropriate gauge and global/bookkeeping charges to the fields which are the basic variables which generate the polynomial ring. In most cases, it is easy to assign the charges for the Cartan generators of Lie groups by looking at the transformation of the fields \footnote{We will discuss in the next section a case of partial gauge fixing which has residual gauge symmetry and the charge assignment will be different.}. These charges are then used as the grading exponents in the multi--graded Hilbert series. More details can be found in the next section \ref{hilbertbh} where we discuss various abelian and non--abelian cases. We set fugacities $\left\{t_{\text {global}}\right\}$ and $\left\{s_{\text {gauge}}\right\}$ for a set of commuting abelian charges under the bookkeeping $U(1)_B$ and the Cartans of $U(N)$ gauge group(s) respectively. We use modified versions of the algorithms available in computer packages such as Macaulay 2\cite{M2} and Singular\cite{Singular} to compute the multi--graded Hilbert series of the quotient ring $R_V$, 
\begin{align}
 H(t ; s_1,\ldots, s_m) & =\sum_{p, i_1,\ldots,i_m} h(p; i_1,\ldots,i_m)\, t^p s_1^{i_1}\ldots s_m^{i_m} \\
 & = \sum_{p, i_1,\ldots,i_m} \chi_{i_1,\ldots,i_m}(s_1,\ldots s_m)\, t^p \, ,  \notag
\end{align}
where $\chi_{i_1,\ldots,i_m}(s_1,\ldots s_m)$ are the characters of some representation (not necessarily irreducible) of the $U(N)$ gauge group(s) .\\
For the purposes of this paper we are not interested in the properties of these characters. We are only interested in extracting the net charge--zero sector from these characters i.e. the part independent of $(s_1,\ldots s_m)$. This will be the ring of gauge invariant operators. We use the \textit{Molien--Weyl formula} \cite{Hanany_adjointsqcd,Pouliot1998MolienFF} to extract the gauge invariant part of the Hilbert series
\begin{equation}
\int \ldots \int d \mu_G(s_1,\ldots s_m)\, H(t ; s_1,\ldots s_m)=H_{G I}(t) \, ,
\end{equation}
where $d \mu_G(s_1,\ldots s_m)$ is the \textit{Haar measure} on the gauge group(s), such that \footnote{We have $U(N)$ groups which can be written as $U(N) \simeq U(1) \times SU(N)$. The expressions for Haar measures for $SU(N)$ groups can be found here \cite{Hanany_adjointsqcd}.}
\begin{equation} \label{haarmeasures}
\int d\mu_G=\begin{cases}
\displaystyle
\int_{U(1)} d \mu_{U(1)}(s) \rightarrow \frac{1}{2 \pi i} \oint_{|s|=1} \frac{d s}{s} &\text {$:U(1)$} \vspace{0.25cm}\\ \displaystyle \int_{SU(2)} d \mu_{SU(2)}(s) \rightarrow \frac{1}{4 \pi i} \oint_{|s|=1} \frac{d s}{s}\left(1-s^2\right)\left(1-s^{-2}\right) & \text{$:SU(2)$}\end{cases}
\end{equation}
For zero dimensional varieties, $H_{GI}(1)=\text{\# of  points}=\text{\# of BH microstates}$.

\section{Counting Black Holes as Hilbert series} \label{hilbertbh}
In this section we will construct the Hilbert series for the abelian and a few non--abelian charge configurations. The computational complexity of Hilbert series is similar to that of finding the Gröbner basis. The number of variables and their corresponding F--term equations scales as $3 N^2 + 6 N+ 15$ for charge configurations of the type $(1,1,1,N)$. So, even after decent speedup due to modifications in existing algorithms and parallelizing the processes whenever possible, we were limited by our modest computational resources \footnote{We are using a 56 core 128 GB machine but the algorithms tend to store history of a large number of matrix manipulations which quickly exhaust the RAM.}. \\
In the Hilbert series computations, all the $\Phi$ and $Z$ fields carry unit charge under a bookkeeping abelian group $U(1)_B$.  In the language of computational algebraic geometry, the bookkeeping charges form the `heft vector' of the polynomial ring. We calculate the Hilbert series with different degrees of gauge fixing:
\begin{itemize}
    \item Full gauge fixing.
    \item Partial gauge fixing \footnote{Charges under the residual gauge symmetries need to be assigned properly. See sections \ref{abelian} and \ref{nonab}.}.
    \item No gauge fixing.
\end{itemize}
Cases for which we could do all three of them with the current computational capacity, they all agree.

\subsection{Abelian case} \label{abelian}
The abelian case has charge configuration  $N_1=1$, $N_2=1$, $N_3=1$ and $N_4=1$. The relative gauge symmetry is the complexified version of $U(1)_{12} \times U(1)_{23} \times U(1)_{34}$ with fugacities $T_0$, $T_1$ and $T_2$ respectively. $T_3$ is the fugacity for the bookkeeping  $U(1)_B$.
\subsubsection*{Full gauge fixing}
We can gauge fix the system by choosing $Z^{(12)}=1$, $Z^{(23)}=1$ and $Z^{(14)}=1$ keeping an overall diagonal $U(1)$ charge arbitrary. This implies that the all relative $U(1)$ charges are now set to zero. We now construct the multigraded polynomial ring with $Z$ and $\Phi$ fields as variables, each carrying charges/weights ${(0,0,0,1)}$. The corresponding Hilbert series points to a zero dimensional variety,
\begin{equation} \label{sol12gf}
 H(T_{3}) =1+7\,T_{3}+4\,T_{3}^{2} \, ,
\end{equation}
where substituting $T_{3}=1$ gives $H(1)=12$. Thus we have a vacua manifold made up of 12 isolated points which are the 12 supersymmetric ground states. The result matches with that reported in section \ref{algbericbh} and the result from the U--dual system of appendix \ref{14helicitytrace}. 

\subsubsection*{Partial gauge fixing}
In situations where one would like the Hilbert series to keep track of gauge invariant operators of subgroups of the full group, we have to resort to partial gauge fixing. In our case we can 
\begin{itemize}
    \item gauge fix $Z^{(34)}=1$. The corresponding charges are in Table \ref{1111partial1} of appendix \ref{gaugecharges}. The Hilbert series after performing the contour integration for the residual gauge group $U(1)_{12} \times U(1)_{23}$ is 
    \begin{equation}
   H(T_{3})= T_3^3+6 T_3^2+4 T_3+1 \, ,
\end{equation}
where note that the coefficients of various powers of $T_3$ are different from \eqref{sol12gf}; yet $H(1)=12\,$.
\item gauge fix $Z^{(12)}=1$ and $Z^{(23)}=1$. The corresponding charges are in Table \ref{1111partial2} of appendix \ref{gaugecharges}. Again, the Hilbert series gives $H(1)=12\,$.
\end{itemize}

\subsubsection*{No gauge fixing}
We now compute the Hilbert series without any gauge fixing. As discussed earlier, for zero dimensional varieties, the total number of GIOs is the degree of the variety and count the number of supersymmetric ground states.  The gauge charges of various fields are in Table \ref{1111wogfcharges} of appendix \ref{gaugecharges} which can be used to construct the multi--graded Hilbert series $H(T_{0},T_{1},T_{2},T_{3})$. The details of the contour integrations are in appendix \ref{detailedhilbert1111}.  The final gauge invariant Hilbert series is 
\begin{equation}
    H(T_{3})=\left(T_3+1\right){}^2 \left(2 T_3+1\right)
\end{equation}
and yet again $H(1)=12$.  

\subsection{Non--abelian cases} \label{nonab}
For non--abelian cases, the complex scalars $\Phi_i^{(k)}$ are $N_k \times N_k$ hermitian matrices transforming in the adjoint representation of $U(N_k)$. On the other hand, the complex scalar $Z^{(k \ell)}$ becomes a $N_k \times N_{\ell}$ complex matrix transforming in the $\left(N_k, \bar{N}_{\ell}\right)$ representation of $U\left(N_k\right) \times U\left(N_{\ell}\right)$. With this information,  we can assign gauge charges to our fields and construct the multi--graded polynomial ring and compute the corresponding Hilbert series. Here, we discuss two cases $N_1=1$, $N_2=1$, $N_3=1$, $N_4=2$ and $N_1=1$, $N_2=1$, $N_3=1$, $N_4=3$ and the previously unreported case $N_1=1$,  $N_2=1$, $N_3=1$, $N_4=4$. \\
For the charges $(1,1,1,2)$ we computed the Hilbert series with three different degrees of complexified gauge fixing. Due to limited resources, we could compute the Hilbert series of charges  $(1,1,1,3)$ and $(1,1,1,4)$ by doing full complexified gauge fixing only.

\subsubsection{Charge $(1,1,1,2)$}
Here we get 56 as the degree of the zero dimensional variety which matches with the results calculated in section \ref{algbericbh} and the result from the U--dual system \cite{n8Shih:2005qf}.
\subsubsection*{Full gauge fixing}
The relative gauge symmetry is the complexified version of $U(1) \times U(1) \times U(2)$. To completely fix the gauge, we set $Z^{(12)}=1$, $Z^{(23)}=1$, $Z^{(14)}=\left(\begin{array}{ll}
1 & 0
\end{array}\right)$ 
and 
$Z^{(24)}=\left(\begin{array}{ll}
0 & 1
\end{array}\right)$. All the fields carry charge $(0,0,0,0,1)$, last one being the bookkeeping charge. The Hilbert series is
\begin{equation} \label{sol56gf1}
 H(T_{4}) =1+16\,T_{4}+39\,T_{4}^{2} \,,
\end{equation}
 where $T_{4}$ is the fugacity corresponding to the bookkeeping charge and $H(1)=56\,$.

 \subsubsection*{Partial gauge fixing}
Various combinations of partial gauge fixing can be performed but here we highlight the case of fixing one of the $U(1)$ and the $U(2)$ gauge symmetry such that the \textit{residual gauge symmetry} induce charges to the fixed gauge groups. The fields under the complexified gauge group transform as 
\begin{equation}\label{eq:gt}
\begin{array}{r}
Z^{(k \ell)} \rightarrow a_k\left(a_{\ell}\right)^{-1} Z^{(k \ell)}, \quad Z^{(4 k)} \rightarrow\left(a_k\right)^{-1} M Z^{(4 k)}, \quad Z^{(k 4)} \rightarrow a_k Z^{(k 4)} M^{-1} \\
\text { for } \quad 1 \leq k \leq 3, \quad k \neq \ell, \\
\Phi_i^{(k)} \rightarrow \Phi_i^{(k)}, \quad \Phi_i^{(4)} \rightarrow M \Phi_i^{(4)} M^{-1}, \quad \text { for } \quad 1 \leq i \leq 3, \quad 1 \leq k \leq 3,
\end{array}
\end{equation}
where $a_k$ for $1 \leq k \leq 3$ are complex numbers and $M$ is a $2 \times 2$ complex matrix. We gauge fix $Z^{(23)}=1$, $Z^{(14)}=\left(\begin{array}{ll}
1 & 0
\end{array}\right)$ 
and 
$Z^{(24)}=\left(\begin{array}{ll}
0 & 1
\end{array}\right)$ leaving behind a residual gauge symmetry \footnote{For the purpose of assigning charges let $U(1)_1=e^{ia}, \, U(1)_2=e^{ib},\, U(1)_3 =e^{ic}$ and $ \displaystyle U(2)=M= \left(\begin{array}{cc}
p & q \\
r & s
\end{array}\right)\,$. To get the residual gauge symmetry, we have 
\begin{itemize}
    \item from invariance of $Z^{(14)}$, $e^{ia} \left(\begin{array}{ll} 
1 & 0
\end{array}\right) M^{-1} =\left(\begin{array}{ll} 
1 & 0
\end{array}\right)\,$. 
\item from invariance of $Z^{(24)}$, $e^{ib} \left(\begin{array}{ll} 
0 & 1
\end{array}\right) M^{-1} =\left(\begin{array}{ll} 
0 & 1
\end{array}\right)\,$.
\item from invariance of $Z^{(23)}$, $1=e^{ib} \, 1 \, e^{-ic}\,$.
\end{itemize}
Therefore, the residual gauge group is $\displaystyle M=\left(\begin{array}{cc}
e^{ia} & 0 \\
0 & e^{ib}
\end{array}\right)$ and $e^{ib}=e^{ic}$ and we will use two diagonal $U(1)_a$ and $U(1)_b$ of $M$ to assign the charges.}. The charge assignment is given in Table \ref{1112partialz} and Table \ref{1112partialphi} of appendix \ref{gaugecharges} with fugacities $T_{0}$, $T_{1}$, $T_{2}$, $T_{3}$, $T_{4}$ and $T_{5}$. After computing the the multi--graded Hilbert series we have to set $T_{3}=T_{0}$, $T_{4}=T_{1}$ and $T_{2}=T_{1}$ as per our gauge choice. The overall $U(1)$ drops out at the end and finally we get $H(1)=56\, $.

\subsubsection*{No gauge fixing}
The relative gauge symmetry is the complexified version of $U(1) \times U(1) \times U(2) \equiv U(1) \times U(1) \times U(1) \times SU(2)$. The charges are mentioned in Table \ref{1112wogfz} and Table \ref{1112wogfphi} of appendix \ref{gaugecharges} where 2 and $-2$ are the adjoint $SU(2)$ gauge charges. For doing the contour integrals we have used the $SU(2)$ Haar measure mentioned in \eqref{haarmeasures}. The answer again is $H(1)=56\,$.

\subsubsection{Charge $(1,1,1,3)$}
Here we get 208 as the degree of the zero dimensional variety which matches with the results discussed in section \ref{algbericbh} and the result from the U--dual system \cite{n8Shih:2005qf}.
\subsubsection*{Full gauge fixing}
In this case, the relative gauge symmetry is the complexified version of $U(1) \times U(1) \times U(3)$. We fix the gauge completely by setting  $Z^{(12)}=1$, $Z^{(23)}=1$, $Z^{(14)}=\left(\begin{array}{lll}
1 & 0 & 0
\end{array}\right)$,  
$Z^{(24)}=\left(\begin{array}{lll}
0 & 1 & 0
\end{array}\right)$
and 
$Z^{(34)}=\left(\begin{array}{lll}
0 & 0 & 1 
\end{array}\right)$ \footnote{An alternate gauge fixing would be to set $Z^{(12)}=1$, $Z^{(23)}=1$, $Z^{(14)}=\left(\begin{array}{lll}
1 & 0 & 0
\end{array}\right)$,  
$Z^{(24)}=\left(\begin{array}{lll}
0 & 1 & 0
\end{array}\right)$
and to choose the first row of any of the $\Phi^{(4)}_i$ field to be $\left(\begin{array}{lll}
a_1 & a_2 & a_3
\end{array}\right)$
where $a_1$, $a_2$ and  $a_3 \neq 0$ are random complex numbers. For arbitrary choices discrete residual gauge symmetries may remain.}.
For both of these gauge fixing, we have the Hilbert series with gauge charges set to $(0,0,0,0,1)$ for all fields. The final Hilbert series is gives  $H(1)=208\,$.

\subsubsection{Charge $(1,1,1,4)$}
Here we get 684 as the degree of the zero dimensional variety which matches with the result from the U--dual system\cite{n8Shih:2005qf}. 

\subsubsection*{Full gauge fixing}
In this case, the relative gauge symmetry is the complexified version of $U(1) \times U(1) \times U(4)$. We can gauge fix it by setting, $Z^{(12)}=1$, $Z^{(23)}=1$, $Z^{(14)}=\left(\begin{array}{llll}
1 & 0 & 0 & 0
\end{array}\right)$,
$Z^{(24)}=\left(\begin{array}{llll}
0 & 1 & 0 & 0
\end{array}\right)$ and 
$Z^{(34)}=\left(\begin{array}{llll}
0 & 0 & 1 & 0
\end{array}\right)$. As we don't have enough $Z$'s to fix the gauge completely, we also set 
$\left(\begin{array}{llll}
(\Phi^{(4)}_{3})_{11} & (\Phi^{(4)}_{3})_{12} & (\Phi^{(4)}_{3})_{13} & (\Phi^{(4)}_{3})_{14}
\end{array}\right)=\left(\begin{array}{llll}
0 & 0 & 0 & 1
\end{array}\right)\,$ \footnote{There are a few caveats in the gauge choices we have been making for the non--abelian cases. For example, we are throwing away the possibility that $Z^{(14)}$ and $Z^{(24)}$ could be parallel to each other or say $\left(\begin{array}{llll}
(\Phi^{(4)}_{3})_{11} & (\Phi^{(4)}_{3})_{12} & (\Phi^{(4)}_{3})_{13} & (\Phi^{(4)}_{3})_{14}
\end{array}\right)=\left(\begin{array}{llll}
0 & 0 & 0 & 0
\end{array}\right)$. We have accounted for these cases but they don't contribute.}. Finally, from the Hilbert series we get $H(1)=684\,$.\\
In principle, we don't foresee any conceptual or technical hurdle to extend these ideas to more general charges $(1,\, 1,\,1\,,N)$, except that we have practical limitations regarding computing resources.

\section{Higher Order Terms in the Potential} \label{quarticz}

In the Lagrangian for the pure D--brane system we presented in section \ref{lagrangian} we have accounted for cubic interactions involving a closed string connecting three branes \eqref{eq:cubic}. It is natural to wonder if we could have quartic interaction term involving fours branes, like 
\begin{equation}\label{quarticzterm}
\mathcal{W}_5=\sqrt{2} C_{4} \sum_{\substack{k,\ell, m, n \\ k< \ell, m,n\\ \ell \neq m \neq n}}^4 \text{(sign)} \, \operatorname{Tr} \left[ Z^{(k \ell)} Z^{(\ell m)} Z^{(m n)} Z^{(n k)}\right]\, .
\end{equation}
The general expectation with Witten index is that it will remain invariant as we change the parameters of the superpotential. For Landau--Ginsberg like models, there are two possibilities as we change the parameters:
\begin{itemize}
    \item Separate critical points collide to form one or more degenerate critical points.
    \item One or more critical points slide towards the far end i.e. infinity. 
\end{itemize}
As an illustration, we take a toy version of the Landau--Ginsberg model discussed in \eqref{landauginzberg} with one hyper--multiplet and a holomorphic superpotential $\mathcal{W}(z)$ such that 
\begin{equation}
    \frac{\partial \mathcal{W}(z)}{\partial z} = \lambda z^2 + 2 z +1 \, ,
\end{equation}
where $\lambda $ is a real tunable parameter. For a moderate range of values for $\lambda$ we get two critical points 
\begin{equation}
    z_1=\frac{\sqrt{1-\lambda }+1}{\lambda } \quad \text{and}\quad z_2= \frac{\sqrt{1-\lambda }-1}{\lambda}
\end{equation}
and the Witten index is $2$. Now, lets take $\lambda$ to the extremes, 
\begin{itemize}
    \item For $\lambda \rightarrow \infty$, we have both $z_1 \rightarrow 0$ and $z_2 \rightarrow 0$\,.
    \item For $\lambda \rightarrow 0$, we have $z_1 \approx -\frac{2}{\lambda }+\frac{1}{2}+\frac{\lambda }{8}+\frac{\lambda ^2}{16}+O\left(\lambda
   ^3\right)$ and $z_2\approx -\frac{1}{2}-\frac{\lambda }{8}-\frac{\lambda ^2}{16}+O\left(\lambda ^3\right)$\,.
\end{itemize}
In the first case, the two separate critical points becomes doubly degenerate and the contribution of the critical point $z=0$ to the Witten index should be counted with its multiplicity, here it is $2$. On the other hand, in the second case the critical point $z_1 $ moves to infinity, and the corresponding quantum SUSY ground state is now outside the Hilbert space. The point $z_2$ remains a valid critical point and the Witten index jumps down from $2$ to $1$. \\
We witness similar behaviour in our system of D--branes if we add the extra quartic term \eqref{quarticzterm}. For concreteness, lets focus on the abelian case where the system has D--brane charges set as $(1,1,1,1)$. In the previous work of one of the current authors \cite{Chowdhury:2014yca,Chowdhury:2015gbk}, looking at the F--term equations \eqref{eq:fterm1}, \eqref{eq:fterm2}, it had been argued that if we parametrically scale down the the coefficients $c^{(k\ell)}$'s by multiplying $\lambda \ll 1$, then the values of the $Z^{(k\ell)}$'s and $\Phi^{(k)}_i$ would scale down as $\sqrt{\lambda}$ and hence the cubic interaction term $\mathcal{W}_2$ \eqref{eq:cubic} dominates over the quartic term $\mathcal{W}_5$ \eqref{quarticzterm}. In this regime, if we include the quartic term, the number of critical points jumps from 12 to 24. Interestingly, out of these 24 critical points, 12 are the critical points of the potential with the original cubic interaction but corrected by $\mathcal{O}(\lambda)$ terms. The other 12 solutions have either some $Z^{(k\ell)}$ or $\Phi^{(k)}_i$ parametrically larger $\displaystyle \mathcal{O}\left(\frac{1}{\lambda^\#}\right)$ than the rest of the numbers. So, it seems Witten index jumps from 12 to 24, spoiling the microstate counting. Same behaviour can also be seen for the non--abelian cases with the addition of many more critical points if quartic superpotential is included in the potential.\\
We will now argue that in both the abelian and non--abelian cases, without changing the field content of the low energy Lagrangian, no such extra superpotential terms survive the constraints coming from various exchange symmetries of type IIA string theory on $T^{6}$ as mentioned in section \ref{symmetries}. To begin with, if we  take the cubic term $\mathcal{W}_2$ and assign arbitrary coefficients $\alpha_i$'s to its individual distinct terms, 
then the invariance of the Lagrangian under the three exchange symmetries constrains the $\alpha_i$'s to specific $\pm \alpha$ and the value of $\alpha$ can be determined by performing a first principle three point vertex amplitude calculation on the open--disk \cite{Chowdhury:2015gbk}. Similarly, $\mathcal{W}_3$ can also be constrained. Coming back to $\mathcal{W}_5$, if we assign arbitrary coefficients as in the case for $\mathcal{W}_2$ and demand invariance under the exchange symmetries, then all such coefficients take zero value i.e. $\mathcal{W}_5$ ceases to exist.  
Other higher order terms which schematically can be written as  $\Phi \Phi Z Z$, $\Phi Z Z Z$, $ (ZZ) (ZZ)$, $ (Z Z Z) (Z Z)$ etc. are also absent for not being compatible with the three exchange symmetries. All these cases and more have been checked both for abelian and non--abelian cases. This ensures that the Witten index is now protected which in turn protects the microstate counting. 

\section{Discussion} \label{discussion}

In this paper we have presented a bouquet of techniques coming from computational algebraic geometry to probe the SUSY vacua manifold of black holes in $\mathcal{N}=8$ string theory compactified on $T^6$. These $\frac{1}{8}$--BPS black holes have a pure D--brane, D2--D2--D2--D6 configuration wrapping various cycles of $T^6$ and admits an U--dual description in terms of a D1--D5--P--KK monopole system \cite{n8Shih:2005qf}. In previous works of one of the authors \cite{Chowdhury:2014yca,Chowdhury:2015gbk}, it has been shown that for very low lying charges the U--duality remains intact and for generic choices of the metric and $B$--field moduli the SUSY vacua manifold presents itself as a collection of isolated points carrying zero angular momentum. This is consistent with the near horizon $AdS_2 \times S^2$ geometry of extremal black holes and puts strong constraints on the fuzzball program \cite{Heidmann2018TheFS,Bena2018AdS2HM}. It is therefore important that we extend the validity of this zero angular  momentum conjecture to more charge configurations. \\
Our primary motivation to take up this work is to eventually count black holes in $\mathcal{N}=2$ theories which are compactified on Calabi-Yau spaces \cite{Greene1997StringTO,n2deBoer:2008zn,n2Manschot:2011xc}. As these manifolds don't have non--trivial 1--cycles, the corresponding black holes can't admit electric or winding charges associated with the $S^1$. Therefore, black holes in Calabi--Yau spaces only carry pure D--brane charges. Currently, we are pushing the counting for $\mathcal{N}=8$ theories with the hope that once enough data (count) is collected, we would be able to write a generating function for different charges and compare with the the U--dual side. Next steps would be to do twining and twisting ($\mathcal{N}=4$ theory on K3 \cite{n4Dijkgraaf:1996it, n4Shih:2005uc, n4Gaiotto:2005hc}) of the theory and perform the counting. We are actively pursuing these directions and hopefully will report new results soon.\\
The results presented in \cite{Chowdhury:2014yca,Chowdhury:2015gbk} are still at the limit of what the latest version of Mathematica can do in terms of solving a system of polynomial equations -- the F--term equations enumerating the isolated SUSY ground states. In our quest to break past the glass ceiling of computational complexity, we ended up discovering a number of very interesting and efficient ways to probe algebraic varieties, especially the zero--dimensional varieties \cite{Mehta2012NumericalAG,Sturmfels1998PolynomialEA}. The techniques discussed in this paper have a far wider scope than black hole physics. Polynomial systems describe a wide variety of systems in physics \cite{Grf2022HilbertST,yanghuiheNumericalEA}, chemistry \cite{chemDickenstein2015BIOCHEMICALRN,chemfaulstich2023algebraic}, mathematics \cite{cox1}, biology \cite{bioGross2015NumericalAG,bioHuggins2006TowardsTH}, engineering \cite{en+2007,en10.1115/1.1649965}, machine learning \cite{aiKnoll2016FixedPO,ml9394420,mlDouglas2021FromAG}, economics \cite{ecMCKELVEY1997411} among other fields of study. We hope that other researchers in our community would take up these techniques and with suitable additions and(or) modifications apply to other interesting settings. \\
For mathematicians, one can take a slightly different view of our non--abelian system for general charges $(N_1, N_2, N_3, N_4)$. The F--terms can now be thought of as a 4--parameter family of sparse polynomial systems where unlike most cases in literature where the number of equations and monomials don't change with parameters, here the number of equations scale like $\mathcal{O}(N_1^2N_2^2N_3^2N_4^2)$ and for each equation the number of monomials at max scale as $\mathcal{O}(N_1+N_2+N_3+N_4)$. This provides an infinite class of sufficiently complex  polynomial families with both discrete and continuous symmetries which can be used to benchmark current and future techniques \& algorithms arising in computational algebraic geometry. They have a nice advantage over other such large polynomial systems as the numbers of solutions for all charges is  known and admits a generating function, a weak Jacobi form of weight $-2$ and index 1 coming from the U--dual description of the system, see Appendix \ref{14helicitytrace}.  We would like to explore if specialized techniques could be developed for such scaling behaviour as many quiver gauge theories share similar scaling \cite{Douglas1996DbranesQA}. The problem can also be recast in the language of optimization theory, where a cost function, the potential $V_F$ needs minimization. Our current endeavours are at the moment a bit handicapped due to lack of access to a state of the art computing facility but hopefully, it will get resolved in future. \\
We are currently looking into the case of setting some of the 
 moduli $c^{(k \ell)}=0$. The $c^{(k \ell)}$'s depend on the background metric and $B$--field moduli and have been assigned non--zero generic values in current and previous works \cite{Chowdhury:2014yca,Chowdhury:2015gbk}. If we set one or more of them to zero i.e. move towards special points in the moduli space, the solutions hit a \textit{degeneration limit} i.e. some fields go to infinity. We hope to regularize these infinities and apply the techniques developed in this paper to carry out the counting \footnote{As we are moving away from a generic point in the moduli space, it is possible that the vacua manifold ceases to be a collection of points. In such cases we may have to  modify the Hilbert series to keep track of an extra fugacity for a R--charge (angular momentum) and extract the \textit{Betti numbers} and then the \textit{E\"{u}ler number}. We are currently pursuing this direction.}. This will settle the counting of \textit{twined} index for these black holes. We are also looking into the counting for charges $(1,1,N_3,N_4)$ and the three charge models $(N_1\,, N_2\,, N_3)$ and their U--dual descriptions.

\bigskip

\noindent {\bf Acknowledgements:} We wish to thank Ashoke Sen for encouraging us to look beyond Mathematica and suggesting the twined case. A.C. would like to thank Swapnamay Mondal for introducing him to the works of Hanany \& friends. The work of A.C. is supported by IIT Bhubaneswar Seed Grant SP--103. The work of S.M. is supported by fellowship from CSIR, Govt. of India.

\appendix
    
\section{Hilbert series of charge $(1,1,1,1)$} \label{detailedhilbert1111}
In this appendix we shall discuss in detail the procedure leading up to the gauge--invariant Hilbert series for the abelian charges. For the (1,1,1,1) charge system, the relative gauge symmetry is $U(1) \times U(1) \times U(1)$. Consider, $T_0$, $T_1$ and $T_2$ as the fugacities corresponding to the relative $U(1)$'s, i.e. $U(1)_{12}$, $U(1)_{23}$ and $U(1)_{34}$ respectively. The fugacity, $T_3$ corresponds to the bookkeeping $U(1)_B$.  The charges are given in Table \ref{1111wogfcharges} and we choose `degree--reverse--lexicographic' monomial ordering   for fast computation of the multi--graded Hilbert series. The Hilbert series is computed using Macaulay2 \cite{M2}, which has the structure
\begin{equation}
    H(T_0, T_1, T_2, T_3)= \frac{N(T_0, T_1, T_2, T_3)}{D(T_0, T_1, T_2, T_3)}
\end{equation}
where $N(T_0, T_1, T_2, T_3)$ is a very long numerator which we will not present here and $D(T_0, T_1, T_2, T_3)$ is the denominator 
\begin{align}
    D(T_0, T_1, T_2, T_3)= & \left(1-T_3\right){}^6 \left(1-\frac{T_3}{T_0}\right) \left(1-T_0 T_3\right) \left(1-\frac{T_3}{T_1}\right) \left(1-\frac{T_3}{T_0 T_1}\right) \left(1-T_1 T_3\right) \notag \\ & \left(1-T_0 T_1 T_3\right)  \left(1-\frac{T_3}{T_2}\right) \left(1-\frac{T_3}{T_1 T_2}\right) \left(1-\frac{T_3}{T_0 T_1 T_2}\right) \left(1-T_2 T_3\right) \notag \\&\left(1-T_1 T_2 T_3\right) \left(1-T_0 T_1 T_2 T_3\right).
\end{align}
After simplification and appropriate scaling the final denominator is  
\begin{equation} \label{den}
\begin{aligned}
   D(T_0, T_1, T_2, T_3)= & T_0^4 T_1^8 T_2^4 \left(T_0-T_3\right) \left(T_1-T_3\right) \left(T_0 T_1-T_3\right) \left(T_2-T_3\right) \left(T_1 T_2-T_3\right) \\ &\left(T_0 T_1 T_2-T_3\right)  \left(T_0 T_3-1\right) \left(T_1 T_3-1\right) \left(T_0 T_1 T_3-1\right) \left(T_2 T_3-1\right) \\ &\left(T_1 T_2 T_3-1\right) \left(T_0 T_1 T_2 T_3-1\right)\,.
\end{aligned}
\end{equation}
Now, we shall extract the count of independent gauge--invariant operators from the  Hilbert series i.e. the part independent of $T_0, T_1, T_2$. To that end, we have to perform contour integrals over the poles of these the variables i.e.
\begin{equation}
H(T_3)=\int  \int \int d \mu_{{U(1)}_{12}} d \mu_{{U(1)}_{23}} d \mu_{{U(1)}_{34}} H(T_0, T_1, T_2, T_3) \,.
\end{equation}
Here, the Haar measure for the $U(1)$ is given by,
\begin{equation}
\int_{\mathrm{U}(1)} \mathrm{d} \mu_{\mathrm{U}(1)}(T) = \frac{1}{2 \pi i} \oint_{|T|=1} \frac{\mathrm{d} T}{T}\,.
\end{equation}
Therefore, the integral we have to perform is 
\begin{equation}
H(T_3)= \frac{1}{(2 \pi i)^3} \oint \frac{d T_2}{T_2}  \oint \frac{d T_1}{T_1} \oint \frac{d T_0}{T_0} \, \frac{N(T_0, T_1, T_2, T_3)}{D(T_0, T_1, T_2, T_3)} \,.
\end{equation}
To perform the integrals we choose circles as contours with an ordering of radii as, $T_3 < T_2 < T_1 < T_0$ \footnote{It can be shown that any other ordering gives the same result.}. To do the $T_0$ integral we choose the contour $|T_0|=1$ and look at its poles at, 
\begin{equation}
T_0 = 0 , \,T_3, \frac{T_3}{T_1}, \,\frac{T_3}{T_1 T_2}, \,\frac{1}{T_3}
,\, \frac{1}{T_1 T_3} \quad \text{and} \quad \frac{1}{T_1 T_2 T_3} \,.    
\end{equation} 
The Hilbert series should be expandable as  power series in terms of the fugacities which puts a \textit{convergence condition} on the terms in the denominator, 
\begin{align} \label{convergence}
& \left| \frac{T_3}{T_0}\right| <1, \, \left| \frac{T_3}{T_0 T_1}\right| <1, \,\left| \frac{T_3}{T_0 T_1 T_2}\right| <1,\,\left| T_3\right| <1,\left| \frac{T_3}{T_1}\right| <1,\, \left| \frac{T_3}{T_2}\right| <1,\,\left| \frac{T_3}{T_1 T_2}\right| <1, \, \left| T_0 T_3\right| <1, \, \notag \\
& \left| T_1 T_3\right| <1, \left| T_0 T_1 T_3\right| <1, \left| T_2 T_3\right| <1,\left| T_1 T_2 T_3\right| <1 \quad \text{and} \quad \left| T_0 T_1 T_2 T_3\right| <1 \, .
\end{align}
 From the above condition we see that the poles at $ \displaystyle T_0 =  0, \, T_3,\, \frac{T_3}{T_1}, \, \frac{T_3}{T_1 T_2}$ are inside the contour $|T_0|=1$ and the rest are to be neglected. After performing the integration, we get the Hilbert series $H(T_1,T_2,T_3)$. Similarly, we perform the $|T_1|=1$ contour integration followed by the $|T_2|=1$ integration to get the gauge--invariant Hilbert series as,
\begin{equation}
    H(T_{3})=\left(T_3+1\right){}^2 \left(2 T_3+1\right) \, .
\end{equation}
We can see that $H(1)=12$ which is the correct count for microstates for the abelian case. Similar calculations follow for the non--abelian cases with appropriate Haar measures.
\section{The $14^{th}$ Helicity Trace Index} \label{14helicitytrace}

In this appendix, we will discuss the $14^{th}$ helicity trace index in the context of $\frac{1}{8}$--BPS, 4--charge black holes.  Following the notation of \cite{Sen:entropyfunction}, and by appropriate duality transformation as described in \cite{Chowdhury:2014yca}, we can write the electric $Q$ and magnetic charge vector $P$ as,
\begin{equation}
Q=\left(\begin{array}{c}
0 \\
-N_2 \\
0 \\
-N_1
\end{array}\right), \quad P=\left(\begin{array}{c}
N_3 \\
0 \\
N_4 \\
0
\end{array}\right)\, .
\end{equation}
The T--duality invariant combinations of charges are
\begin{equation}
Q^2=2 N_1 N_2, \quad P^2=2 N_3 N_4, \quad Q \cdot P=0\,  .
\end{equation}
With the following definition of `gcd'
\begin{equation}
\begin{aligned}
& \ell_1=\operatorname{gcd}\left\{Q_i P_j-Q_j P_i\right\}=\operatorname{gcd}\left\{N_1 N_3, N_1 N_4, N_2 N_3, N_2 N_4\right\}, \\
& \ell_2=\operatorname{gcd}\left\{Q^2 / 2, P^2 / 2, Q \cdot P\right\}=\operatorname{gcd}\left\{N_1 N_2, N_3 N_4\right\} \, ,
\end{aligned}
\end{equation}
the $14^{th}$ helicity super--trace $B_{14}$, can be written as \cite{n8Sen:2009gy}
\begin{equation}
B_{14}=- \sum_{s \mid \ell_1 \ell_2} s\, \hat{c}\left(\Delta / s^2\right), \quad \Delta \equiv Q^2 P^2=4 N_1 N_2 N_3 N_4\,,
\end{equation}
where $\hat{c}(u)$ is defined through the relation \cite{n8Shih:2005qf,Maldacena1999CountingBB,Pioline:2005vi}
\begin{equation}
-\vartheta_1(z \mid \tau)^2\, \eta(\tau)^{-6} \equiv \sum_{k, l} \widehat{c}\left(4 k-l^2\right) e^{2 \pi i(k \tau+l z)} \,.
\end{equation}
$\vartheta_1(z \mid \tau)$ and $\eta(\tau)$ are respectively the odd Jacobi theta function and the Dedekind eta function. \\
Table \ref{B14} shows helicity trace index for different 4--charges relevant for this paper. 
\begin{table}[h!]
    \centering
    $\begin{array}{|c|c|c|c|c|}
   \hline  \text{Charges}    &  \operatorname{gcd}\left\{\ell_1, \ell_2\right\}  & s &\Delta & B_{14}\\
   \hline   (1,1,1,1)   & 1 & 1 & 4 &-\widehat{c}(4) = 12\\
   \hline   (1,1,1,2)   & 1 & 1 & 8 &-\widehat{c}(8) = 56\\
   \hline   (1,1,1,3)   & 1 & 1 & 12 &-\widehat{c}(12) = 208\\
   \hline   (1,1,1,4)   & 1 & 1 & 16 &-\widehat{c}(16) = 684\\  
   
   \hline
    \end{array}$
    \caption{$14^{th}$ helicity trace index for different 4--charges.}
    \label{B14}
\end{table}

\section{Gauge charges} \label{gaugecharges}
In this appendix we collect various gauge charge assignment tables arising in section \ref{hilbertbh} discussing Hilbert series. In the tables we have removed the $\Phi$'s which are set to zero to `gauge fix' the phase symmetries.
\begin{table}[H]  
\centering
$\begin{array}{|c||c|c|c|c|c|c|c|c|c|c|c|c|c|c|c|c|c|c|}
 \hline U(1)_{12} & 1 & -1 & 0 & 0 & 0 & 0 & 0 & 0 & 0 & 0 & 0 & 0 & 0 & 0 & 0 & 0 & 0 & 0 \\
\hline U(1)_{23} & 0 & 0 & 1 & -1 & 1 & -1 & 1 & -1 & 1 & -1 & 0 & 0 & 0 & 0 & 0 & 0 & 0 & 0 \\
\hline U(1)_{34} & 0 & 0 & 0 & 0 & 1 & -1 & 0 & 0 & 1 & -1 & 0 & 0 & 0 & 0 & 0 & 0 & 0 & 0 \\
\hline U(1)_{B} & 1 & 1 & 1 & 1 & 1 & 1 & 1 & 1 & 1 & 1 & 1 & 1 & 1 & 1 & 1 & 1 & 1 & 1 \\
\hline
\end{array}
$
\caption{Gauge charges for fields in the order $\{Z^{(12)}, Z^{(21)}, Z^{(13)}, Z^{(31)}, Z^{(14)}, Z^{(41)}, \\ Z^{(23)},  Z^{(32)},  Z^{(24)}, Z^{(42)}, Z^{(34)}, 
Z^{(43)}, \Phi^{(2)}_{3}, \Phi^{(3)}_{1}, \Phi^{(3)}_{2}, \Phi^{(4)}_{1}, \Phi^{(4)}_{2}, \Phi^{(4)}_{3}\}$.} \label{1111partial1}
\end{table}
\begin{table}[h!]
\centering
$\begin{array}{|c||c|c|c|c|c|c|c|c|c|c|c|c|c|c|c|c|c|c|}
\hline U(1)_{12} &0 & 0 & 0 & 0 & 0 & 0 & 0 & 0 & 0 & 0 & 0 & 0 & 0 & 0 & 0 & 0 & 0 & 0  \\
\hline U(1)_{23} &0 & 0 & 0 & 0 & 0 & 0 & 0 & 0 & 0 & 0 & 0 & 0 & 0 & 0 & 0 & 0 & 0 & 0  \\
\hline U(1)_{34} &0 & 0 & 0 & 0 & 1 & -1 & 0 & 0 & 1 & -1 & 1 & -1 & 0 & 0 & 0 & 0 & 0 & 0 \\
\hline U(1)_{B} &1 & 1 & 1 & 1 & 1 & 1 & 1 & 1 & 1 & 1 & 1 & 1 & 1 & 1 & 1 & 1 & 1 & 1  \\
\hline
\end{array}
$ 
\caption{Gauge charges for fields in the order $\{Z^{(12)}, Z^{(21)}, Z^{(13)}, Z^{(31)}, Z^{(14)}, Z^{(41)}, \\ Z^{(23)},  Z^{(32)},  Z^{(24)}, Z^{(42)}, Z^{(34)}, 
Z^{(43)}, \Phi^{(2)}_{3}, \Phi^{(3)}_{1}, \Phi^{(3)}_{2}, \Phi^{(4)}_{1}, \Phi^{(4)}_{2}, \Phi^{(4)}_{3}\}$.} \label{1111partial2}
\end{table}
\begin{table}[h!]
\centering
$\begin{array}{|c||c|c|c|c|c|c|c|c|c|c|c|c|c|c|c|c|c|c|} 
\hline U(1)_{12} & 1 & -1 & 1 & -1 & 1 & -1 & 0 & 0 & 0 & 0 & 0 & 0 & 0 & 0 & 0 & 0 & 0 & 0  \\
\hline U(1)_{23} & 0 & 0 & 1 & -1 & 1 & -1 & 1 & -1 & 1 & -1 & 0 & 0 & 0 & 0 & 0 & 0 & 0 & 0  \\
\hline U(1)_{34} & 0 & 0 & 0 & 0 & 1 & -1 & 0 & 0 & 1 & -1 & 1 & -1 & 0 & 0 & 0 & 0 & 0 & 0  \\
\hline U(1)_{B} & 1 & 1 & 1 & 1 & 1 & 1 & 1 & 1 & 1 & 1 & 1 & 1 & 1 & 1 & 1 & 1 & 1 & 1 \\
\hline
\end{array}$ \\
\caption{Gauge charges for fields in the order $\{Z^{(12)}, Z^{(21)}, Z^{(13)}, Z^{(31)}, Z^{(14)}, Z^{(41)}, \\ Z^{(23)}, Z^{(32)}, Z^{(24)}, Z^{(42)}, Z^{(34)}, 
Z^{(43)}, \Phi^{(2)}_{3},  \Phi^{(3)}_{1}, \Phi^{(3)}_{2},  \Phi^{(4)}_{1}, \Phi^{(4)}_{2}, \Phi^{(4)}_{3}\}$. 
} \label{1111wogfcharges}
\end{table}
\begin{table}[h!] \small
\centering
$\begin{array}{|c||c|c|c|c|c|c|c|c|c|c|c|c|c|c|c|c|c|c|}
\hline U(1)_{1}& 1 & -1 & 1 & -1 & 0 & 0 & 1 & 1 & -1 & -1 & 0 & 0 & 0 & 0 & 0 & 0 & 0 & 0 \\
\hline U(1)_{2}& -1 & 1 & 0 & 0 & 1 & -1 & 0 & 0 & 0 & 0 & 1 & 1 & -1 & -1 & 0 & 0 & 0 & 0 \\
\hline U(1)_{3}& 0 & 0 & -1 & 1 & -1 & 1 & 0 & 0 & 0 & 0 & 0 & 0 & 0 & 0 & 1 & 1 & -1 & -1 \\
 \hline U(1)_{a}& 0 & 0 & 0 & 0 & 0 & 0 & -1 & 0 & 1 & 0 & -1 & 0 & 1 & 0 & -1 & 0 & 1 & 0 \\
\hline U(1)_{b} & 0 & 0 & 0 & 0 & 0 & 0 & 0 & -1 & 0 & 1 & 0 & -1 & 0 & 1 & 0 & -1 & 0 & 1 \\
\hline U(1)_{B} & 1 & 1 & 1 & 1 & 1 & 1 & 1 & 1 & 1 & 1 & 1 & 1 & 1 & 1 & 1 & 1 & 1 & 1 \\
\hline
\end{array}
$
\caption{Gauge charges for $Z^{(k\ell)}$ fields in the order $\{Z^{(12)}, Z^{(21)}, Z^{(13)}, Z^{(31)}, Z^{(23)},\\ Z^{(32)}, Z^{(14)}_{1}, Z^{(14)}_{2}, Z^{(41)}_{1}, Z^{(41)}_{2}, Z^{(24)}_{1}, Z^{(24)}_{2}, Z^{(42)}_{1}, Z^{(42)}_{2}, Z^{(34)}_{1},Z^{(34)}_{2}, Z^{(43)}_{1}, Z^{(43)}_{2}\}$.} \label{1112partialz}
\end{table}
\begin{table}[h!]
\centering
$\begin{array}{|c||c|c|c|c|c|c|c|c|c|c|c|c|c|c|c|}
\hline U(1)_{1} & 0 & 0 & 0 & 0 & 0 & 0 & 0 & 0 & 0 & 0 & 0 & 0 & 0 & 0 & 0 \\
\hline U(1)_{2} & 0 & 0 & 0 & 0 & 0 & 0 & 0 & 0 & 0 & 0 & 0 & 0 & 0 & 0 & 0 \\
\hline U(1)_{3}  & 0 & 0 & 0 & 0 & 0 & 0 & 0 & 0 & 0 & 0 & 0 & 0 & 0 & 0 & 0 \\
\hline U(1)_{a} & 0 & 0 & 0 & 0 & 1 & -1 & 0 & 0 & 1 & -1 & 0 & 0 & 1 & -1 & 0 \\
\hline U(1)_{b}  & 0 & 0 & 0 & 0 & -1 & 1 & 0 & 0 & -1 & 1 & 0 & 0 & -1 & 1 & 0 \\
\hline U(1)_{B}  & 1 & 1 & 1 & 1 & 1 & 1 & 1 & 1 & 1 & 1 & 1 & 1 & 1 & 1 & 1 \\
\hline
\end{array}
 $ \\ 
\caption{Gauge charges for $\Phi^{(k)}_{\ell}$ fields in the order $\{\Phi^{(2)}_{3}, \Phi^{(3)}_{1},  \Phi^{(3)}_{2}, \Phi^{(4)}_{1,11}, \Phi^{(4)}_{1,12}, \Phi^{(4)}_{1,21},\\  \Phi^{(4)}_{1,22}, \Phi^{(4)}_{2,11}, \Phi^{(4)}_{2,12}, \Phi^{(4)}_{2,21}, \Phi^{(4)}_{2,22}, 
 \Phi^{(4)}_{3,11}, \Phi^{(4)}_{3,12}, \Phi^{(4)}_{3,21}, \Phi^{(4)}_{3,22}\}$.} \label{1112partialphi}
 \end{table}
\begin{table}[h!]
\centering
$\begin{array}{|c||c|c|c|c|c|c|c|c|c|c|c|c|c|c|c|c|c|c|}
\hline U(1)_{12}& 1 & -1 & 1 & -1 & 0 & 0 & 1 & 1 & -1 & -1 & 0 & 0 & 0 & 0 & 0 & 0 & 0 & 0 \\
\hline U(1)_{23}& 0 & 0 & 1 & -1 & 1 & -1 & 1 & 1 & -1 & -1 & 1 & 1 & -1 & -1 & 0 & 0 & 0 & 0 \\
\hline U(1)_{34}& 0 & 0 & 0 & 0 & 0 & 0 & 1 & 1 & -1 & -1 & 1 & 1 & -1 & -1 & 1 & 1 & -1 & -1 \\
\hline SU(2)& 0 & 0 & 0 & 0 & 0 & 0 & 1 & -1 & -1 & 1 & 1 & -1 & -1 & 1 & 1 & -1 & -1 & 1 \\
\hline U(1)_{B}& 1 & 1 & 1 & 1 & 1 & 1 & 1 & 1 & 1 & 1 & 1 & 1 & 1 & 1 & 1 & 1 & 1 & 1 \\
\hline
\end{array}
$
\caption{Gauge charges for $Z^{(k\ell)}$ fields in the order $\{Z^{(12)}, Z^{(21)}, Z^{(13)}, Z^{(31)}, Z^{(23)}, \\ Z^{(32)}, Z^{(14)}_{1}, Z^{(14)}_{2}, Z^{(41)}_{1}, Z^{(41)}_{2}, Z^{(24)}_{1}, Z^{(24)}_{2}, Z^{(42)}_{1}, Z^{(42)}_{2}, Z^{(34)}_{1}, Z^{(34)}_{2}, Z^{(43)}_{1}, Z^{(43)}_{2}\}$.} \label{1112wogfz} 
\end{table}
\begin{table}[H]
\centering
$\begin{array}{|c||c|c|c|c|c|c|c|c|c|c|c|c|c|c|c|}
\hline U(1)_{12}  & 0 & 0 & 0 & 0 & 0 & 0 & 0 & 0 & 0 & 0 & 0 & 0 & 0 & 0 & 0 \\
\hline U(1) _{23} & 0 & 0 & 0 & 0 & 0 & 0 & 0 & 0 & 0 & 0 & 0 & 0 & 0 & 0 & 0 \\
\hline U(1)_{34}& 0 & 0 & 0 & 0 & 0 & 0 & 0 & 0 & 0 & 0 & 0 & 0 & 0 & 0 & 0 \\
\hline SU(2) & 0 & 0 & 0 & 0 & -2 & 2 & 0 & 0 & -2 & 2 & 0 & 0 & -2 & 2 & 0 \\
 \hline U(1)_{B}  & 1 & 1 & 1 & 1 & 1 & 1 & 1 & 1 & 1 & 1 & 1 & 1 & 1 & 1 & 1 \\
 \hline
\end{array}
$ \\ 
\caption{Gauge charges for $\Phi^{(k)}_{\ell}$ fields in the order $\{ \Phi^{(2)}_{3},  \Phi^{(3)}_{1}, \Phi^{(3)}_{2}, \Phi^{(4)}_{1,11}, \Phi^{(4)}_{1,12}, \Phi^{(4)}_{1,21}, \\ \Phi^{(4)}_{1,22},  \Phi^{(4)}_{2,11}, \Phi^{(4)}_{2,12}, \Phi^{(4)}_{2,21}, \Phi^{(4)}_{2,22},
 \Phi^{(4)}_{3,11}, \Phi^{(4)}_{3,12}, \Phi^{(4)}_{3,21}, \Phi^{(4)}_{3,22}\}$.} \label{1112wogfphi}
\end{table}


\bibliographystyle{unsrt}
\bibliography{references} 

\end{document}